\documentclass[a4paper,fleqn]{cas-sc}

\usepackage[authoryear]{natbib}
\usepackage{graphicx}
\usepackage{subcaption}
\usepackage{adjustbox}
\usepackage{float}
\usepackage{amsmath, amssymb, amsfonts}
\usepackage{gensymb}
\def\tsc#1{\csdef{#1}{\textsc{\lowercase{#1}}\xspace}}
\tsc{WGM}
\tsc{QE}

\begin{document}
\let\WriteBookmarks\relax
\def\floatpagepagefraction{1}
\def\textpagefraction{.001}

\shorttitle{}    

\shortauthors{}  

\title [mode = title]{Comparing statistical learning models in
wastewater-based epidemiology: An application to
norovirus}  



%

\author[1]{Caelan McNamara}[orcid = 0009-0007-0080-7057]
\fnmark[1]
\credit{Methodology, Writing - Original Draft}
\author[1]{Fion Tan}
\fnmark[1]
\credit{Methodology, Writing - Original Draft, Visualisation}
\author[1,2]{Ella White}[orcid= 0009-0002-0739-1094]
\ead{ella.white19@imperial.ac.uk}
\fnmark[1]
\cormark[1]\cortext[1]{Corresponding author}
\credit{Conceptualization of this study, Methodology, Supervision, Data Curation, Writing - Review \& Editing, Visualisation}
\author[1,3]{Elizaveta Semenova}[orcid= 0000-0002-8271-2575]
\credit{Methodology, Supervision, Writing - Review \& Editing}
\author[1,2]{Marta Blangiardo}[orcid=0000-0002-1621-704X]
\credit{Conceptualization of this study, Methodology, Supervision, Writing - Review \& Editing}
\fntext[1]{These authors contributed equally to this work}

\affiliation[1]{organization={School of Public Health, Imperial College London},
            addressline={90 Wood lane}, 
            city={London},
            postcode={W12 0BZ}, 
            country={UK}}
            
\affiliation[2]{organization={MRC Centre for Environment and Health, Imperial College London},
            addressline={White City Campus, Wood lane}, 
            city={London},
            postcode={W12 0BZ}, 
            country={UK}}

\affiliation[3]{organization={MRC Centre for Global Infectious Disease Analysis, Imperial College London},
            addressline={White City Campus, Wood lane}, 
            city={London},
            postcode={W12 0BZ}, 
            country={UK}}


\begin{abstract}
Wastewater-based epidemiology (WBE) is an increasingly important tool for infectious disease surveillance, but there has been limited direct comparison of modelling approaches for predicting pathogen concentrations across space and time. We compare the predictive performance of six modelling approaches using norovirus in England as a case study, considering 3,232 wastewater samples from 152 sewage treatment works collected between May 2021 and March 2022 as part of the UK Environmental Monitoring for Health Protection programme. The benchmark was a Bayesian spatio-temporal model using Integrated Nested Laplace Approximation (INLA) with the Stochastic Partial Differential Equation approach (SPDE), compared with Lasso regression, Generalised Additive Models (GAM), Bayesian GAM, Extreme Gradient Boosting (XGBoost), and Random Forest. Models were evaluated using 10-fold spatial-block cross-validation with metrics including mean squared error, bias, correlation, empirical coverage of 95\% prediction intervals, interval score, and computational cost. Random Forest was the best overall performing model, achieving the best interval score and nearest to nominal empirical coverage (95.3\%), while maintaining point prediction accuracy comparable to XGBoost. The INLA-SPDE model also performed well, with near-nominal empirical coverage, the third best interval score, and consistently low bias across all evaluated metrics. Comparing predicted spatio-temporal trends, both models yielded similar spatial patterns but notable differences in uncertainty estimation. Our findings highlight a trade-off between predictive accuracy, uncertainty quantification, and computational efficiency. Ensemble machine learning methods are well suited to rapid prediction, whereas Bayesian geostatistical models remain valuable when probabilistic decision support is a priority for public health surveillance.
\end{abstract}


\begin{highlights}
\item First paper to compare several modelling approaches for spatio-temporal data in the context of wastewater-based epidemiology
\item Random Forest best predicts norovirus concentrations across wastewater sites 
\item Bayesian models offer more principled uncertainty quantification of the predictions
\item Machine learning and statistical models show clear accuracy-uncertainty trade-offs
\end{highlights}


\maketitle

\section{Introduction}\label{sec:intro}

Wastewater-based epidemiology (WBE) has emerged as a valuable tool for monitoring infectious diseases at the population level. By analysing viral concentrations in wastewater, WBE provides a community-wide measure of disease burden that is less sensitive to biases in clinical testing and healthcare-seeking behaviour \citep{Gholipour2022OccurrenceReview, Bowes2024WBE}. The approach has been used to detect and predict outbreaks of pathogens including norovirus and hepatitis A \citep{Hellmer2014DetectionOutbreaks}, and gained particular prominence during and after the COVID-19 pandemic to complement conventional clinical surveillance \citep{Walker2024PilotingEngland, Li2023APandemic}. Because it captures shedding from both symptomatic and asymptomatic infections, WBE is especially valuable for population-level public-health surveillance \citep{Bowes2024WBE}.

A range of statistical and machine learning methods have been applied to WBE data. Early studies often relied on relatively simple regression approaches, but larger surveillance networks require models that can account for spatial and temporal dependence. Li et al.\ (2023) developed a Bayesian spatio-temporal framework for predicting viral concentrations over a continuous spatial domain while quantifying uncertainty \citep{Li2023APandemic}. Machine learning methods have also been used in this setting, including Random Forest and Extreme Gradient Boosting (XGBoost) models linking SARS-CoV-2 wastewater concentrations to case counts \citep{Koureas2021WastewaterMunicipalities, Mills2024ThePrevalence}. In the context of norovirus surveillance, research has compared machine learning algorithms for norovirus outbreak classification and showed that Support Vector Machines and Random Forest performed well \citep{Cho2024ImprovingMethods}. Other statistical learning approaches such as Generalised Additive Models (GAMs) have likewise been used to capture non-linear relationships in WBE and broader epidemiological applications \citep{Cabrera2019ModellingModels}. Despite the predictive performance of machine learning methods, they often lack interpretability and native uncertainty quantification, both of which are important for public-health decision-making \citep{Ennab2022DesigningHealthcare}.

Despite increasing use of these methods in WBE, there has been no thorough direct comparison between Bayesian spatio-temporal models and commonly used statistical learning approaches. In this study, using national-scale wastewater monitoring data for norovirus in England, we compare a Bayesian spatio-temporal model with five comparator approaches: Lasso regression, GAM, Bayesian GAM, XGBoost, and Random Forest. We adopt the framework proposed by \citet{Li2023APandemic} as the benchmark model and evaluate predictive accuracy and computational cost across methods. We then discuss the implications of these trade-offs for the design of wastewater-based public-health surveillance.

We focus on norovirus as a particularly suitable case study for WBE. Norovirus is a leading cause of acute gastroenteritis worldwide and is associated with a substantial health and economic burden in England \citep{Patel2008SystematicGastroenteritis, Tam2016EconomicKingdom, Harris2017Re-assessingPopulation, NationalGOV.UK}. Surveillance based on reported cases substantially under-ascertains true community transmission because many infections are asymptomatic or never present to clinical care \citep{Qi2018GlobalMeta-analysis, Wang2023GlobalMeta-analysis, Ondrikova2023ComparisonSurveillance, NationalGOV.UK}. This makes norovirus well suited to wastewater surveillance, which can capture both symptomatic and asymptomatic shedding at population scale and relatively low cost \citep{Gholipour2022OccurrenceReview, Walker2024PilotingEngland}. 

The remainder of the paper is structured as follows: Section \ref{sec:Methods} presents the wastewater data and covariates used in the analysis, the spatial-block cross-validation design, introduces the different modelling approaches and the uncertainty quantification. Section \ref{sec:results} presents the results including predictive performance and spatio-temporal trends. Section \ref{sec:discussion} discusses the main findings, limitations, and implications for wastewater-based public-health surveillance.

\section{Methods}\label{sec:Methods}

\subsection{Data}
\label{sec:data}

\begin{figure}[pos=h]
    \centering
\includegraphics[width=0.4\textwidth]{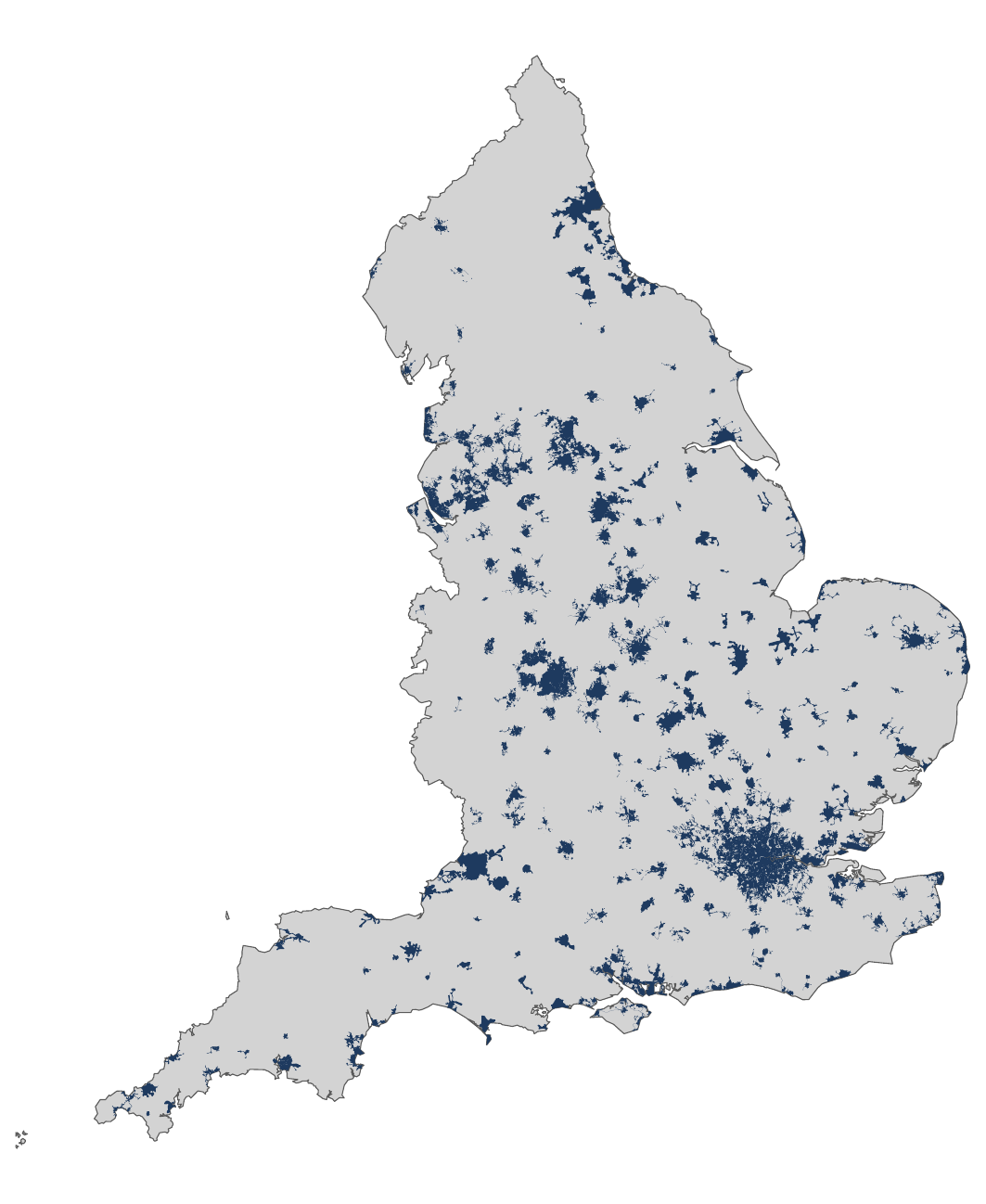}
    \caption{Catchment areas of the 152 sewage treatment works (STWs) included in the study. Polygons represent the spatial extent of each STW catchment used for wastewater monitoring.}
    \label{fig:stw_catchment}
\end{figure}

Wastewater samples and associated metadata were collected as part of the Environmental Monitoring for Health Protection (EMHP) programme, which was established in 2020 to monitor SARS-CoV-2 and other priority pathogens in wastewater across England \citep{Wade2022UnderstandingProgrammes}. The COVID-19 component of the programme concluded in March 2022, although the underlying sampling infrastructure has informed subsequent national wastewater monitoring activity led by the UK Health Security Agency (UKHSA). In this study, we analysed norovirus concentrations from 3{,}232 samples collected at 152 sewage treatment works (STWs) across England between 27 May 2021 and 30 March 2022, with each site sampled approximately fortnightly. Observations were aggregated to weekly resolution, yielding a temporal index $t = 1, \ldots, 45$ over the study period; not every site contributed an observation in every week. A map of the STW catchment areas across England is shown in Figure \ref{fig:stw_catchment}. Norovirus concentration measurements were population and flow normalised, available as log$_{10}$ gene copies per 100, 000 capita, as described in \cite{roberts_data_2022}. Raw wastewater data were obtained from the Centre for Environment, Fisheries and Aquaculture Science (CEFAS) Data Portal \citep{CefasView}.

A range of demographic and environmental covariates were included as predictors of norovirus concentrations. Demographic indicators were obtained from the 2021 Census at lower super output area (LSOA) level \citep{CensusStatistics}. The proportion of the population identifying as Black, Asian and Minority Ethnic (BAME), included as a proxy for differential exposure and household composition which has been linked to norovirus infection susceptibility \citep{Tsang2018TransmissibilityChina}. The Index of Multiple Deprivation (IMD) was included to capture dimensions of socioeconomic deprivation including income, employment, health, education, crime, housing access, and living environment \citep{CensusStatistics}.  LSOA to STW catchment mapping of IMD and BAME was performed using the population-weighted approach outlined by \citet{Li2023APandemic}. This was using the STW-LSOA lookup table provided by \citet{hoffmann_wastewater_2022}.

School and care home coordinate locations were obtained from UK Government open data \citep{DownloadsGOV.UK, CareData.gov.uk} and converted to densities per 100{,}000 population at LSOA and STW catchment level. These were included because schools and care homes cluster vulnerable individuals and are well-documented sites of norovirus outbreaks, with prolonged shedding occurring in older adults and young children \cite{Simmons2013DurationCDC, Lai2013AAge, Lian2019Epidemiology20142017}. Land cover characteristics were summarised using the proportion of urban land cover from the CORINE dataset at 100\,m resolution \cite{CORINEService}, included as a proxy for population structure and greater urbanicity has been associated with increased infectious disease transmission \citep{Alirol2011Urbanisation}. The proportion of agricultural land cover was also included as a potential 
confounder, as agricultural run-off from nitrogen-based fertlisers can elevate ammonia concentrations in wastewater influent 
\citep{Fowler2013TheNitrogen}. Since ammonia is used as a proxy for flow volume at sites where 
direct flow measurement is unavailable, artificially elevated ammonia 
concentrations may confound the flow-normalisation process \citep{Walker2024PilotingEngland, roberts_data_2022}.

Meteorological covariates were obtained from the ERA5 Climate Data Store  \citep{ERA5Present} at approximately 9\,km resolution (0.25$\degree$ latitude-longitude grid) , including  total precipitation and air temperature. Norovirus outbreaks show seasonality and have been linked to temperature and humidity \citep{Bruggink2010TheRainfall, Lopman2009HostWales}. Rainfall was additionally included as heavy precipitation events can increase surface water ingress and cause transient dilution of viral concentrations in wastewater that may not be fully captured by 
flow-normalisation \citep{Wade2022UnderstandingProgrammes}.  Hourly values were aggregated to daily which was then processed as a 7 day rolling average.  These values were aligned to STW and LSOA boundaries using the \texttt{terra} package, which facilitates extraction and area-weighted averaging of gridded raster data to irregular polygon geometries \citep{Hijmans2024terra}. Both rainfall and temperature were categorised into quintiles based on their empirical distributions to accommodate any non-linear relationships with norovirus concentration.

Human mobility was represented using the percentage change in workplace mobility relative to baseline from the Google COVID-19 Community Mobility Reports \citep{COVID-19Reports}, capturing person-to-person contact patterns relevant to norovirus transmission \citep{Arthur2017ContactEcology}. This was available at the local authority district (LAD) 2019 boundaries and downscaled to STW catchment using a population-weighted average of similar to the LSOA-STW catchment mapping used by \citet{Li2023APandemic}. LAD-LSOA mapping was done using the lookup table provided by the Office of National Statistics \citep{ONS2023LookupTable}.

Indicators for COVID-19 policy phases (lockdown step 3, 17 May to 18 July 2021; lockdown step 4, 19 July to 7 December 2021; plan B restrictions, 8 December 2021 to 26 January 2022; post-restrictions, 27 January to 28 March 2022) \citep{TimelineGovernment, Lockdown_2021_2022Storage} were included to account for behavioural effects on transmission since wastewater norovirus signals were observed to fall under restrictions and rebound when restrictions were lifted \citep{Wade2022UnderstandingProgrammes}. Summary statistics for all covariates included in the analysis are presented in Table~\ref{tab:summary_stats}.

\begin{table}[ht]
\centering
\caption{Summary statistics of covariates included in the models}
\begin{adjustbox}{width=\textwidth}  
\begin{tabular}{>{\raggedright\arraybackslash}p{5.5cm}
                >{\centering\arraybackslash}p{2cm}
                >{\centering\arraybackslash}p{2.5cm}
                >{\raggedright\arraybackslash}p{4.5cm}
                >{\centering\arraybackslash}p{2.5cm}}
\toprule
Variable & Missing (\%) & Mean (SD) & Spatial resolution & Temporal resolution \\
\midrule
Log$_{10}$ norovirus concentration (gene copies/ 100k capita) & 53.1 & 5.16 (0.69) & STW  & Weekly \\
Index of Multiple Deprivation  & 0.0 & 23.08 (7.56) & LSOA & N/A \\
Proportion BAME  & 0.0 & 0.14 (0.12) & LSOA & N/A \\
School density (per 100{,}000) & N/A & N/A & Point location & N/A \\
Care home density (per 100{,}000) & N/A & N/A & Point location & N/A \\
Proportion agricultural land & 0.0 & 0.14 (0.13) & 1\,km $\times$ 1\,km grid & N/A \\
Proportion urban land & 0.0 & 0.63 (0.19) & 1\,km $\times$ 1\,km grid & N/A \\
Population mobility (\% change in travel to workplace) & 0 & $-$24.07 (9.32) & Local authority district & Daily \\
Rainfall (mm) & 0 & 8.47 (2.49) & 0.25$\degree$ $\times$ 0.25$\degree$ grid & Hourly \\
Temperature (K) & 0 &  283.79 ( 4.79) & 0.25$\degree$ $\times$ 0.25$\degree$ grid & Hourly \\
\bottomrule
\end{tabular}
\end{adjustbox}
\vspace{0.2cm}
\begin{minipage}{0.85\textwidth}
\footnotesize
\textit{Abbreviations:}STW= sewage treatment work; LSOA= lower super output area; BAME = Black, Asian and Minority Ethnicities; K= kelvin. Missingness for norovirus concentration reflects any STW-week pairing that does not have a sample.
\end{minipage}

\label{tab:summary_stats}
\end{table}

\subsection{Spatial-block Cross-validation}
\label{sec:spatial_block}

Cross-validation was used to compare the approaches presented below. As ordinary cross-validation based on random sampling can produce overly optimistic performance estimates for spatial models because nearby observations tend to be more similar due to spatial autocorrelation \citep{Pohjankukka2017EstimatingValidation, Ploton2020SpatialModels}, we used 10-fold spatial-block cross-validation. The 152 STW sites were partitioned into 10 spatially contiguous blocks using the \texttt{blockCV} package \citep{Valavi2019BlockCV:Models}: a regular grid was overlaid on the study area, sites within the same grid cell were grouped, and grid cells were then assigned to folds with block size chosen to approximately match the empirical spatial range of norovirus concentrations \citep{Ploton2020SpatialModels}. Each fold was held out in turn as the validation set while the remaining nine folds formed the training set, ensuring spatial separation between training and validation data. Final fold sizes and partition are reported in the Appendix (Table~\ref{tab:spatial_folds} and  Figure~\ref{fig:validation_site}).

\subsection{INLA-SPDE Model}
\label{sec:bayesian_model}

\subsubsection{Model specification}
\label{sec:notation}

The first method considered is a  Bayesian geostatistical model, adapted from  the framework outlined in \cite{Li2023APandemic}. The log$_{10}$-transformed norovirus concentration at STW $i = 1, \ldots, 152$ and week $t = 1, \ldots, 45$ is modelled as
\begin{equation}
y_{it} \sim \text{Normal}(\mu_{it}, \sigma_y^2),
\label{eq:y_it}
\end{equation}
where $\sigma_y^2$ is the measurement-error variance. For the latent mean concentration $\mu_{it}$ we specify a linear model
\begin{equation}
\mu_{it} = \alpha + \mathbf{X}_{it}\boldsymbol{\beta} + u_i + v_t + z_{it},
\label{eq:mu_it}
\end{equation}
where $\alpha$ is the overall intercept and $\mathbf{X}_{it} = (x_{1it}, x_{2it}, \ldots, x_{mit})$ is the vector of $m = 9$ covariates introduces in Section~\ref{sec:data}. The term $u_i$ is a catchment-level random effect with exchangeable prior $u_i \sim \text{Normal}(0, \sigma_u^2)$, and $v_t$ is a temporal random effect with exchangeable prior $v_t \sim \text{Normal}(0, \sigma_v^2)$. The term $z_{it}$ is a spatio-temporal interaction that captures local departures from the global spatial and temporal patterns. We assume a separable model for $z_{it}$, with white noise specification across $t$ for each $i$ and, for each time point, a spatial dependence specified through a zero-mean Gaussian field with Matérn covariance:
\begin{equation}
  \text{Cov}(z_{it}, z_{jt}) \;=\; \sigma_z^2 \, \frac{2^{1-\lambda_{\nu}}}{\Gamma(\lambda_{\nu})} \big(\kappa d_{ij}\big)^{\lambda_{\nu}} K_{\lambda_{\nu}}\!\big(\kappa d_{ij}\big)  
\end{equation}

where $\Gamma(\cdot)$ is the gamma function, $K_{\lambda_{\nu}}$ is the modified Bessel function of the second kind of order $\lambda_{\nu} > 0$ (controlling smoothness), and $d_{ij} = \|\mathbf{s}_i - \mathbf{s}_j\|$ is the Euclidean distance between locations $\mathbf{s}_i$ and $\mathbf{s}_j$. We fix the smoothness parameter $\lambda_{\nu} = 1$ to avoid identifiability issues that commonly arise when smoothness is estimated jointly with the range and variance from data of this size. The scale parameter $\kappa > 0$ is linked to the practical range $r = \sqrt{8\lambda_{\nu}}/\kappa$, defined as the distance at which spatial correlation falls to approximately 0.1 \cite{Lindgren2011AnApproach}.

\subsubsection{Prior Specification}
The intercept $\alpha$ and the fixed effects $\boldsymbol{\beta}$ were assigned weakly informative priors $\text{Normal}(0, 10^4)$. Penalised complexity (PC) priors \citep{Simpson2017PenalisingPriors} were used for the random-effect standard deviations and the Matérn hyperparameters. We set $\Pr(\sigma > 10) = 0.05$ on the standard deviations of $u_i$, $v_t$, and the spatio-temporal interaction, expressing the prior belief that these standard deviations are likely below 10 with only a small probability of exceeding it. For the spatial range, we set $\Pr(r < 5) = 0.05$, encoding a weak preference for ranges greater than 5\,km. This prior places negligible probability on ranges below 5 km, which is well supported by the empirical variogram (Appendix \ref{appendix:variogram}) indicating a spatial range of approximately 20 km. The likelihood precision $1/\sigma_y^2$ was assigned a $\text{Gamma}(1, 0.2)$ prior. Further details on the Bayesian spatio-temporal modelling framework are given in \citep{Blangiardo2015SpatialR-INLA}.

\subsubsection{Model Fitting}

\begin{figure}[pos=!h]
  \centering
  \includegraphics[width=0.5\textwidth]{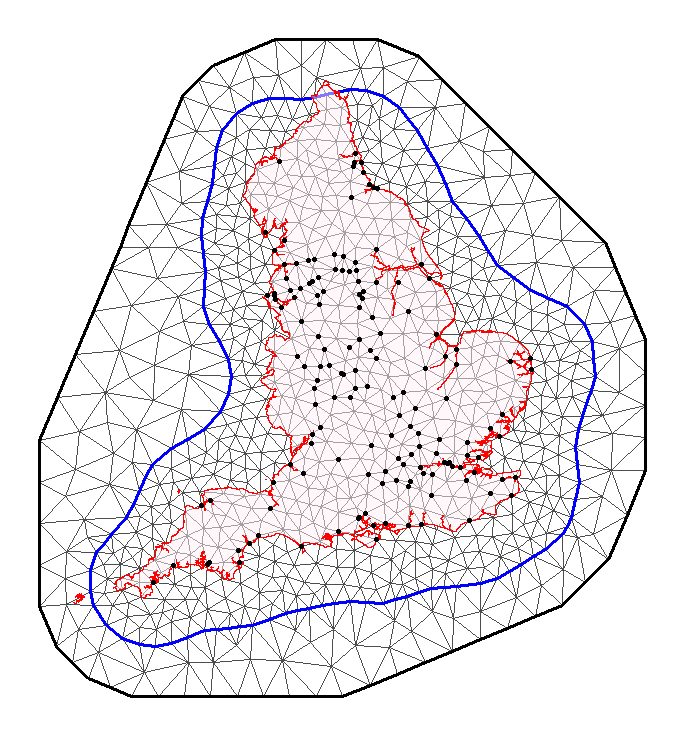}
  \caption{Mesh implemented for model fitting in INLA with England boundary. Solid dots represent the 152 locations of sewage treatment works (STWs) in the study. }
  \label{fig:mesh plot}
\end{figure}

The model was implemented using Integrated Nested Laplace Approximation (INLA) via the \texttt{R-INLA} package to estimate the joint posterior $\pi(\boldsymbol{\Theta} \mid \bf{y})$ where $\boldsymbol{\Theta} =\{\alpha, \boldsymbol{\beta}, \boldsymbol{u}, \boldsymbol{v}, \boldsymbol{z}, \sigma_u^2, \sigma_v^2, \sigma_z^2, \sigma_y^2, r \}$. INLA offers substantial computational gains over Markov chain Monte Carlo (MCMC) while providing accurate marginal posterior approximations \citep{Rue2009ApproximateApproximations}. INLA can be coupled with the Stochastic Partial Differential Equation (SPDE) approach to predict over a continuous spatial domain when data are observed at point locations \citep{Lindgren2011AnApproach}, as in \citet{Li2023APandemic}. The SPDE approach approximates the continuous spatial field using weighted basis functions defined at the vertices of a triangulated mesh over the study region. See Appendix \ref{appendix:spde} for more details. The mesh used here is shown in Figure~\ref{fig:mesh plot}

\subsection{Comparator models}
\label{sec:ML}

We compared the Bayesian spatio-temporal model against a suite of alternative approaches for predicting weekly log$_{10}$-transformed norovirus concentrations. All comparators used the same standardised covariates as in Section~\ref{sec:data} and were evaluated under the same spatial 10-fold cross-validation scheme (Section~\ref{sec:spatial_block}). Implementations for all the approaches were in R (v4.4.2). 

\subsubsection{Lasso Regression}
Lasso regression extends ordinary linear regression with an $L_1$ penalty to induce sparsity and perform automatic variable selection \citep{Friedman2010Glmnet:Models}:
\begin{equation}
\hat{\boldsymbol{\beta}}^{\text{lasso}} = \arg\min_{\boldsymbol{\beta}} \left\{ \frac{1}{n} \sum_{i=1}^{N}\sum_{t=1}^{T} \left( y_{it} - \alpha - \mathbf{X}_{it}\boldsymbol{\beta} \right)^2 + \tau \|\boldsymbol{\beta}\|_1 \right\},
\end{equation}
where $\tau \ge 0$ controls the regularisation strength. The optimal $\tau$ was selected by 10-fold cross-validation. Notation follows Equation \ref{eq:mu_it} in Section~\ref{sec:notation}. The approach was implemented through the \texttt{glmnet} package \citep{Friedman2010Glmnet:Models}.

\subsubsection{Generalised Additive Model (GAM)}
GAMs allow smooth non-linear effects of selected predictors \citep{Wood2001Mgcv:R}:
\begin{equation}
y_{it} = \alpha + \mathbf{X}_{it}\boldsymbol{\beta} + f_s(\mathbf{s}_i) + f_t(t) + \epsilon_{it}, \qquad 
\end{equation}
where $\mathbf{s}_i = (\text{longitude}_i, \text{latitude}_i)$, $f_s(\cdot)$ is a bivariate thin-plate regression spline for spatial location, and $f_t(\cdot)$ is a univariate thin-plate spline for time \citep{Wood2003ThinSplines}. Smoothing parameters and effective degrees of freedom were selected by restricted maximum likelihood (REML). Thin-plate regression splines were chosen for their flexibility in capturing non-linear spatial and temporal variation while controlling for overfitting through the REML penalty.  The approach was implemented through the \texttt{mgcv} package \citep{Wood2017GeneralizedEdition}.

\subsubsection{Bayesian Generalised Additive Model}
\label{sec:bayesian_gam}
The Bayesian GAM extends the GAM framework by placing prior distributions on all model parameters and approximating posterior predictive distributions via Hamiltonian Monte Carlo. The model specification is
\begin{equation}
y_{it} \mid \mu_{it} \sim \mathcal{N}(\mu_{it}, \sigma_y^2)
\end{equation}
as specified in Equation~\ref{eq:y_it}. Priors were chosen to mirror those of the INLA-SPDE model where possible, to support comparability between the two Bayesian frameworks:
\begin{align*}
\alpha &\sim \mathcal{N}(0, 10^4), \\
\beta_k &\sim \mathcal{N}(0, 10^4), \\
\sigma_y &\sim \text{Student-}t(3, 0, 2.5).
\end{align*}
The Student-$t(3, 0, 2.5)$ prior on $\sigma_y$ is the \texttt{brms} default and is weakly informative on the standard-deviation scale, regularising estimates away from implausibly large values while leaving the data to dominate the posterior \citep{Burkner2017Brms:Stan, Gelman2008WeaklyAnalyses}. Smooth terms $f_s$ and $f_t$ were represented by basis expansions with each smooth coefficient assigned a Gaussian prior with mean zero and variance determined by the smoothing penalty.

Posterior inference used four MCMC chains, each with 3000 iterations (1000 warmup, 2000 sampling), and a target acceptance rate of 0.95. These settings were selected after preliminary runs confirmed adequate effective sample sizes (> 400 for all parameters of interest) and $\hat{R} < 1.01$, indicating satisfactory convergence \citep{Vehtari2021Rank}. Convergence was additionally assessed by inspecting trace plots and posterior densities (Appendix \ref{appendix:trace}). We did not perform a formal prior sensitivity analysis, however, the priors used here closely match those of the INLA-SPDE model to support fair comparison between the two Bayesian frameworks.The approach was implemented through the \texttt{brms} package \citep{Burkner2017Brms:Stan}.

\subsubsection{Extreme Gradient Boosting (XGBoost)}
XGBoost fits an additive ensemble of regression trees sequentially to minimise a regularised loss \citep{Chen2016XGBoost:System}:
\begin{equation}
\hat{y}_{it} = \sum_{m=1}^{M} f_m(\mathbf{X}_{it}), \qquad f_m \in \mathcal{F},
\end{equation}
where $f_m$ is the $m$th regression tree and $\mathcal{F}$ is the space of regression trees. At each iteration, $f_m$ minimises
\begin{equation}
\text{Obj} = \sum_{i=1}^{N}\sum_{t=1}^{T} \left( y_{it} - \hat{y}_{it}^{(m-1)} - f_m(\mathbf{X}_{it}) \right)^2 + \Omega(f_m),
\end{equation}
with $\Omega(f_m)$ penalising tree complexity. Notation follows Section~\ref{sec:notation}. Final hyperparameters (Section~\ref{sec:optimisation}) were 300 boosting rounds, maximum tree depth 6, learning rate 0.1, subsample ratio 0.8, and column subsampling 1, with squared-error loss. The approach was implemented through the \texttt{xgboost} package \citep{Chen2016XGBoost:System}.

\subsubsection{Random Forest}
Random Forest aggregates predictions from an ensemble of decorrelated decision trees built on bootstrap samples of the data \citep{Breiman2001RandomForests}:
\begin{equation}
\hat{y}_{it} = \frac{1}{B} \sum_{b=1}^{B} T_b(\mathbf{X}_{it}),
\end{equation}
where $T_b(\cdot)$ is the prediction from the $b$th tree, constructed using a random subset of $m$ predictors at each split. Notation follows Section~\ref{sec:notation}. Final hyperparameters (Section~\ref{sec:optimisation}) were $B = 500$ trees and $m = 3$ variables per split. Bootstrapping observations and subsampling predictors at each split decorrelates the trees, reducing variance compared with single decision trees \cite{Breiman2001RandomForests}. The approach was implemented through the \texttt{ranger} package \citep{Liaw2002Classification}. 

\subsubsection{Hyperparameter Optimisation}
\label{sec:optimisation}
For Random Forest, a grid search was performed over $mtry \in \{2, 3, \ldots, 9\}$ (the full range of available covariates) and $ntree \in \{250, 500, 750, 1000\}$. Each candidate combination was evaluated using 10-fold cross-validation repeated twice, with root mean squared error ($\text{RMSE} = \sqrt{\frac{1}{n}\sum_{i=1}^{n}(y_{it} - \hat{y}_{it})^2}$) as the selection criterion. The optimal configuration was $mtry = 3$, $ntree = 500$, which was used in all subsequent Random Forest analyses.

For XGBoost, an initial grid search via the \texttt{caret} package over learning rate, tree depth, and subsampling parameters did not improve performance over the \texttt{xgboost} defaults under spatial cross-validation. Final hyperparameters were therefore set through targeted preliminary experimentation: 300 boosting rounds, maximum tree depth 6, learning rate 0.1, subsampling ratio 0.8, and column subsampling 1, with squared-error loss and early stopping after 20 rounds without improvement.

\subsection{Uncertainty Quantification}
\label{sec:uncertainty}

Uncertainty quantification methods varied by model. For the Bayesian models 
(INLA-SPDE and Bayesian GAM), uncertainty was quantified directly from the 
posterior predictive distribution, yielding $B=1000$ posterior draws per 
observation from which 95\% credible intervals were computed.

For the GAM, uncertainty was quantified via posterior simulation as described in \citep{Simpson2025gratia}. Although fitted using REML rather than full Bayesian inference, GAMs can be interpreted within an empirical Bayes framework \citep{Wood2017GeneralizedEdition}, in which the 
smoothing penalties act as improper priors on the smooth coefficients. This allows an approximate posterior distribution over the model parameters to be 
constructed, from which $B=1000$ draws were simulated to obtain 95\% 
prediction intervals. 

For Lasso, empirical 95\% prediction intervals were constructed via 
non-parametric bootstrap resampling of the training set ($B=1000$ samples), 
with intervals defined by the 2.5th and 97.5th percentiles of the bootstrap 
distribution. For Random Forest, 95\% prediction intervals were obtained 
using the Quantile Regression Forest framework 
\citep{Meinshausen2006QuantileForests}. Similarly, for XGBoost, quantile 
regression was used to construct 95\% predictive intervals.

It should be noted that these uncertainty quantification approaches are not 
strictly equivalent and comparisons across models should therefore be interpreted 
as suggestive rather than definitive.

\subsection{Evaluation Metrics}
\label{sec:validation}

Each metric was computed within each fold and then averaged across the ten spatial folds to obtain the overall performance for each model. To assess computational cost, total wall-clock time (\textit{hh:mm:ss}) was recorded across all ten folds, including both training and prediction. Predictions were generated at monitoring sites in the validation set for each held-out fold. Lower and upper bounds at the nominal 95\% level were also returned. Predictions were compared to observed log$_{10}$-transformed norovirus concentrations using the metrics below:

\begin{enumerate}

\item Mean squared error (MSE) defined as
\[
\text{MSE} = \frac{1}{n}\sum_{i=1}^{n} (y_i - \hat{y}_i)^2.
\]
\item Prediction bias, which quantifies systematic over- or under-prediction
\[
\text{Bias} = \frac{1}{n}\sum_{i=1}^{n} (\hat{y}_i - y_i).
\]

\item Predictive association using the Pearson correlation coefficient
\[
r = \frac{\sum_{i=1}^{n}(y_i - \bar{y})(\hat{y}_i - \bar{\hat{y}})}{\sqrt{\sum_{i=1}^{n}(y_i - \bar{y})^2}\sqrt{\sum_{i=1}^{n}(\hat{y}_i - \bar{\hat{y}})^2}}.
\]

\item Empirical coverage of the 95\% prediction interval defined as the proportion of observations falling within the predicted bounds $[L_i, U_i]$:
\[
\text{Coverage} = \frac{1}{n}\sum_{i=1}^{n} \mathbb{I}(L_i \le y_i \le U_i),
\]
where $\mathbb{I}(\cdot)$ is the indicator function.

\item The interval score (IS) jointly penalises interval width and coverage 
failures \citep{Gneiting2007}. For a nominal $(1-\alpha)$ prediction interval 
$[L_i, U_i]$ with $\alpha = 0.05$, the interval score for observation $i$ is
\[
\text{IS}_{\alpha}(L_i, U_i, y_i) =
    (U_i - L_i)
    + \frac{2}{\alpha}(L_i - y_i)\,\mathbb{I}(y_i < L_i)
    + \frac{2}{\alpha}(y_i - U_i)\,\mathbb{I}(y_i > U_i),
\]
where $\mathbb{I}(\cdot)$ is the indicator function. The overall interval score 
is then averaged across all observations:
\[
\overline{\text{IS}} = \frac{1}{n}\sum_{i=1}^{n}
    \text{IS}_{\alpha}(L_i, U_i, y_i).
\]
A lower interval score indicates better performance, rewarding sharp 
(narrow) intervals that nonetheless achieve adequate empirical coverage.
\end{enumerate}

\section{Results}
\label{sec:results}

\subsection{Observed spatio-temporal trends}
\begin{figure}[pos= h]
    \centering
    \includegraphics[width=1\textwidth]{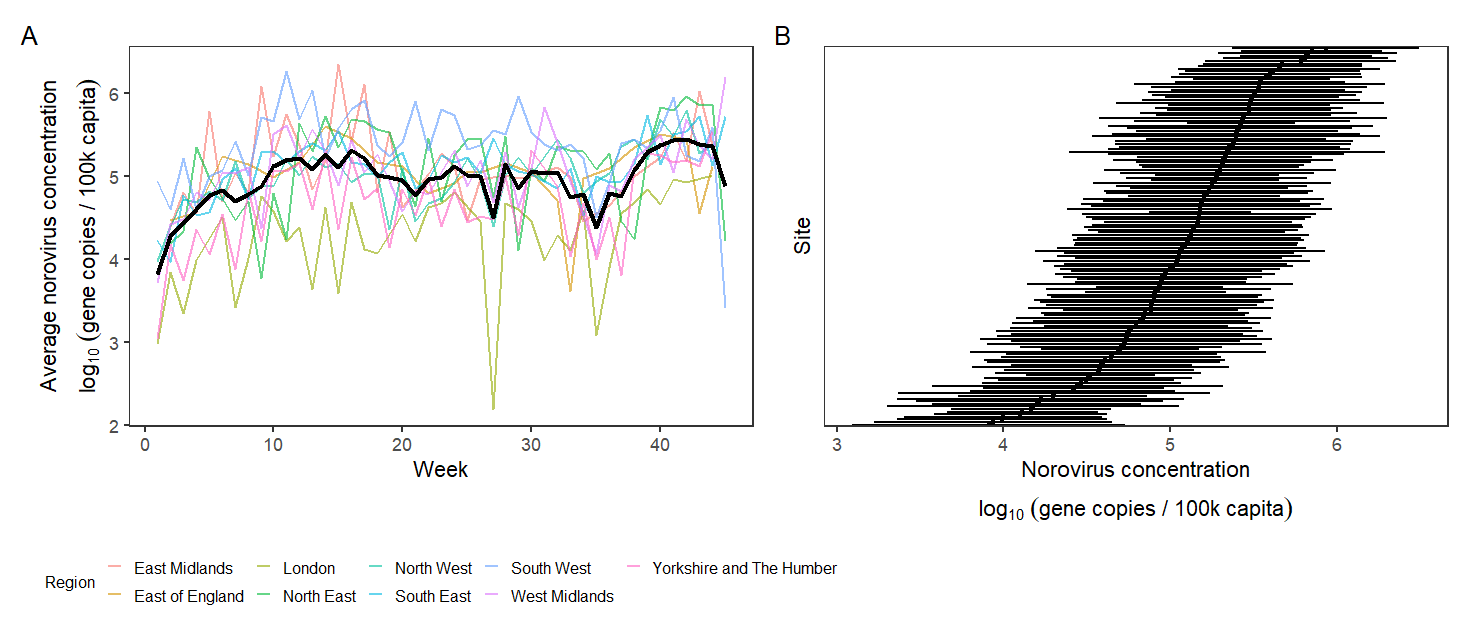}
    \caption{(A) Regional time series of observed mean log$_{10}$ norovirus concentration. National average concentration across all sites is indicated with the black line; (B) Distribution of site-level concentrations with points indicating the mean and error bars indicating  $\pm$1 standard deviation. Concentrations are expressed as gene copies (gc) per 100k capita.}
    \label{fig:norovirus_combined}
\end{figure}

Figure \ref{fig:norovirus_combined} shows the temporal trend of the observed norovirus concentration over the study period at the regional and national level.  Nationally, there are lower values at the start of the study period, then an increase to reach a first peak around August 2021. This was followed by a decrease in January before rising again to a second peak at the end of the study period. At the regional level there is more variation, but all the regions tend to follow the national trend with the exception of London. The distribution of norovirus concentration for each site throughout the study period highlights the overall spatial variation in the norovirus concentrations.

\begin{table}[ht]
 \centering
    \caption{Predictive performance comparison of models. Bold indicates the best value per metric.}\label{tab:results}
    \begin{adjustbox}{max width=\textwidth}
\begin{tabular}{lllllll}
\hline
\textbf{Model} & \textbf{MSE} & \textbf{Bias} & \textbf{Correlation} & \textbf{95 \% Coverage} & \textbf{Interval score} & \textbf{Elapsed time (hh:mm:ss)} \\ \hline
INLA-SPDE      & 0.515        & -0.053        & 0.363                & 93.77                   &  3.580                  & 03:44:41                         \\
Lasso         & 0.504        & \textbf{0.020}         & 0.365                & 13.17                   & 17.77                   & 00:13:43                         \\
GAM            & 0.573        & -0.045        & 0.302                & 92.51                   & 3.77                    & \textbf{00:00:11}                         \\
Bayesian GAM   &   0.534      & -0.0394       &       0.355          &       90.79             &    3.851                &    01:22:56                      \\
Random Forest  & 0.488        & -0.032        & 0.393                & \textbf{95.34}                   & \textbf{3.414}                   & 00:00:20                         \\ 
XGBoost        & \textbf{0.477}        & 0.022         & \textbf{0.413}                & 92.32                   & 3.512                   & 00:00:44                         \\
\bottomrule
\end{tabular}
\end{adjustbox}
\vspace{0.2cm}
\begin{minipage}{0.95\textwidth}
\footnotesize
\textit{Abbreviations:} MSE = mean squared error; INLA-SPDE =  Integrated Nested Laplace Approximation with the Stochastic Partial Differential Equation; GAM = Generalised Additive Model; XGBoost = Extreme Gradient Boosting.  Time reports total wall-clock time aggregated across all ten cross-validation folds.
\end{minipage}

\end{table}

\subsection{Predictive performance}
Across all models, predictive accuracy was broadly similar (Table~\ref{tab:results}), with MSE ranging narrowly from 0.477 to 0.573. However, models differed substantially in their calibration of predictive uncertainty and computational demands. Ensemble tree-based methods achieved the strongest point prediction accuracy: XGBoost yielded the lowest MSE (0.477) and the highest correlation with observed concentrations ($r = 0.413$), with Random Forest performing comparably (MSE = 0.488; $r = 0.393$), 
suggesting these methods were particularly effective at capturing non-linear relationships among the covariates. Among the regression-based approaches, Lasso achieved moderate accuracy (MSE = 0.504; $r = 0.365$), while the classical GAM produced the weakest correlation ($r = 0.302$) alongside the highest MSE (0.573). Both Bayesian models were broadly comparable to the regression-based comparators in point accuracy, with INLA-SPDE achieving an MSE of 0.515 and $r = 0.363$. In terms of bias, the Bayesian GAM exhibited the smallest absolute bias ($-0.039$), with Lasso and XGBoost showing the only positive bias (0.020 and 0.022 respectively).

Calibration of predictive uncertainty revealed the main differences across models and is arguably one of the most important metrics in a public health surveillance context where interval estimates inform decision-making. Random Forest achieved near-nominal empirical coverage (95.3\%), closely followed by INLA-SPDE (93.8\%) and GAM (92.5\%) and XGBoost (92.3\%). Notably, both Bayesian models achieved broadly similar empirical coverage: INLA-SPDE yielded 93.8\% and the Bayesian GAM 90.8\%, both substantially closer to the nominal 95\% threshold than most non-Bayesian comparators, reflecting the more principled uncertainty quantification afforded by their respective posterior predictive distributions. The interval score, which jointly penalises poor sharpness and coverage, further confirmed these rankings: Random Forest achieved the best interval score (3.414), followed by 
XGBoost (3.512), with INLA-SPDE achieving the next best interval score (3.580), ahead of all remaining models including the Bayesian GAM (3.851) and GAM (3.770). This indicates that despite its greater computational cost, the INLA-SPDE model produces well-calibrated and comparatively sharp predictive intervals. In contrast, Lasso severely underestimated uncertainty, with a coverage of only 13.2\% and an interval score of 17.77, far exceeding all other models.

Computation time varied by more than an order of magnitude across approaches, from eleven seconds (GAM) to just under four hours (INLA-SPDE). Random Forest was completed in twenty seconds, making it the most computationally efficient model with strong calibration. Importantly, the cost of INLA-SPDE reflects the complexity of the spatio-temporal model rather than Bayesian inference \emph{per se}, and cross-validation was completed in under four hours while retaining formal uncertainty quantification. 

\subsection{Predicted spatio-temporal trends}
An application of INLA-SPDE models is to predict at unobserved locations which follows naturally from the latent Gaussian field. Here we generated weekly predictions at the population-weighted centroids of all 33{,}755 LSOAs in England. For comparison we also obtained predictions from the best-performing comparator model which was identified to be Random Forest. For the INLA-SPDE model, model estimation and prediction is done jointly to fully propagate the uncertainty. The predicted mean norovirus concentration at LSOA centroid $j$ in week $t$ is $\mu_{jt} = \alpha + \mathbf{X}_{jt}\boldsymbol{\beta} + v_t + z_{jt}$
where $\mathbf{X}_{jt}$ is the covariate profile for LSOA $j$ in week $t$, $v_t$ is the temporal random effect, and $z_{jt}$ is obtained by evaluating the SPDE-approximated Gaussian field at the centroid coordinates of LSOA $j$. As LSOA centroids have no associated monitoring site, $u_i$ from Equation~\ref{eq:mu_it} does not contribute a fixed value to the linear predictor but instead estimates between-site variance $\sigma^2_u$ which is marginalised over during posterior predictive sampling. This ensures that site-level variability is reflected in predictive uncertainty. Posterior predictive samples (n=1000) $y'_{jt}$ are drawn from
\begin{equation}
\pi(\boldsymbol{y'} \mid \boldsymbol{y}) = \int p(\boldsymbol{y'} \mid \boldsymbol{\Theta})\, p(\boldsymbol{\Theta} \mid \boldsymbol{y})\, d\boldsymbol{\Theta},
\label{posterior_predictive}
\end{equation}

where $\boldsymbol{\Theta}$ is sampled from the joint posterior distribution.

Random Forest was selected as the comparator model on the basis of its strongest overall performance across accuracy, uncertainty calibration, and computational efficiency (Table~\ref{tab:results}).The predicted mean norovirus concentration at LSOA centroid $j$ in week $t$ for Random Forest was obtained:
\begin{equation}
\hat{y}_{jt} = \frac{1}{B} \sum_{b=1}^{B} T_b(\mathbf{X}_{jt}),
\end{equation}

To facilitate comparison with the INLA-SPDE framework, individual tree predictions are treated as approximate samples from the predictive distribution, analogous to posterior draws, though we note this interpretation is approximate.  In both cases, uncertainty at any spatio-temporal resolution is obtained by summarising the sample distribution after aggregation, rather than aggregating point estimates alone.

\begin{figure}[pos=!h]
\centering\includegraphics[width=0.9\textwidth]{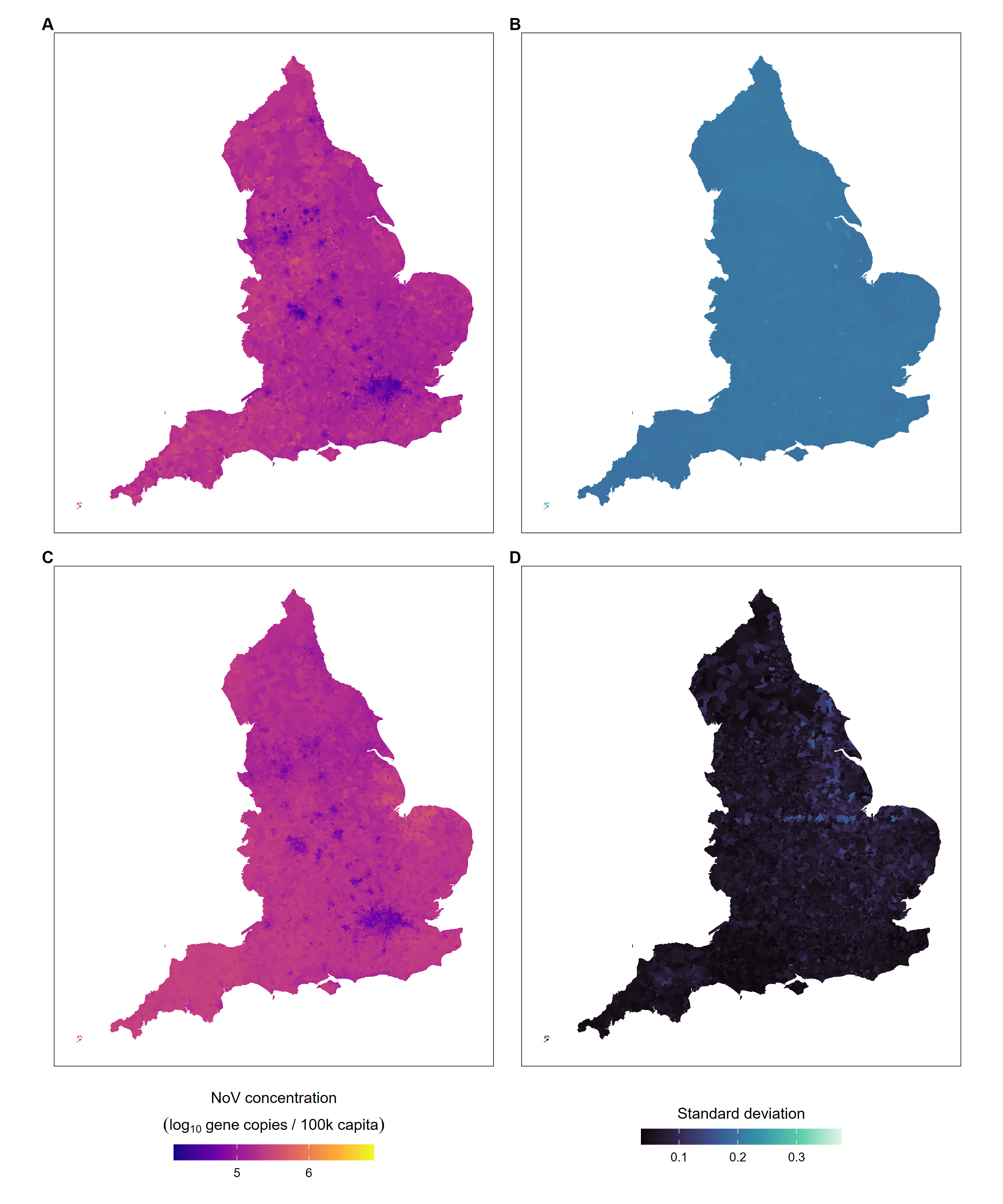}
  \caption{Predicted mean and standard deviation of INLA (top) and Random Forest (bottom) models at LSOA level. (A) Posterior overall mean from INLA model. (B) Posterior standard deviation from INLA model. (C) Overall mean from Random Forest model. (D) Standard deviation from Random Forest Model.}
  \label{fig:spatial_mean}
\end{figure}

Figure \ref{fig:spatial_mean} (A) and (C) visualises the LSOA-level variation in predictions from the INLA-SPDE and Random Forest model respectively. Overall, both models have similar spatial variation in their predictions in their distribution and scale of norovirus concentration. Figure \ref{fig:spatial_mean} (B) and (D) displays the uncertainty of the predictions for the INLA-SPDE and Random Forest models respectively. A more notable difference can be observed here, with the INLA-SPDE model giving an overall higher uncertainty than the Random Forest model. The Random Forest model also exhibits more spatial variation of the uncertainty compared to the INLA-SPDE.

\begin{figure}[pos=!h]
    \centering
    \includegraphics[width=1\linewidth]{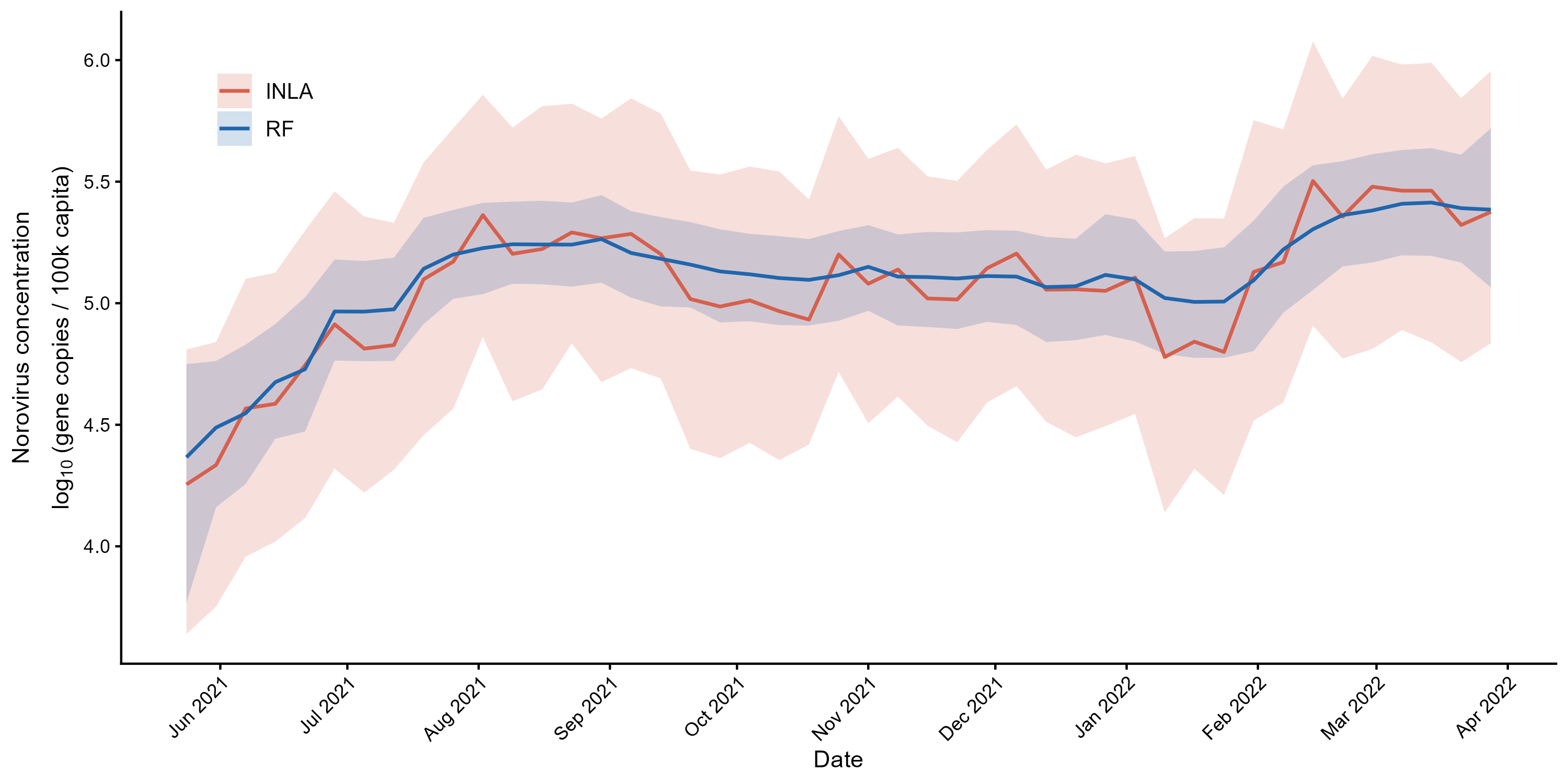}
    \caption{Weekly average predicted norovirus concentration 
 across England with 95 \% predictive intervals from the Random Forest (RF, blue) and INLA-SPDE model (red)}
    \label{fig:nat_pred_comparison}
\end{figure}

Figure \ref{fig:nat_pred_comparison} displays the predicted national trends from the Random Forest and INLA-SPDE models. Both models have consistent trends with the observed national average in Figure \ref{fig:norovirus_combined} in their mean prediction. The Random Forest produced notably narrower intervals and smoother predictions than the INLA-SPDE model.

\clearpage

\begin{figure}[pos=h]
\centering\includegraphics[width=0.9\textwidth]{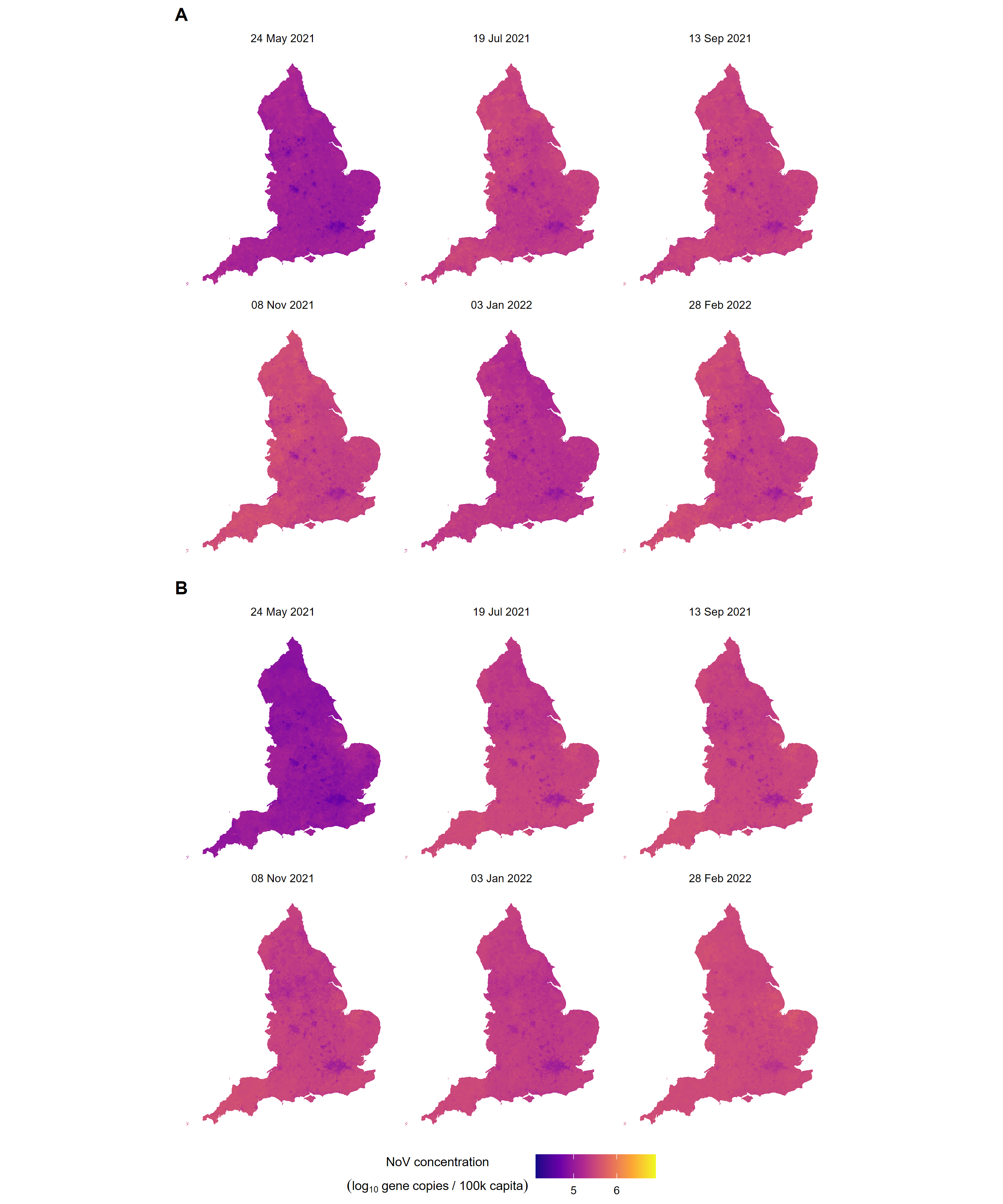}
 \caption{Weekly predictions of norovirus concentrations, showing every 8th week. Panels A and B correspond to the predictions generated by the INLA-SPDE and Random Forest models respectively.}\label{fig:spatiotemporal}
\end{figure}

Figure \ref{fig:spatiotemporal} shows the predicted spatio-temporal trends at the LSOA level for both the INLA-SPDE and Random Forest models. Both models generated similar weekly predictions, with consistent temporal trends as Figure \ref{fig:nat_pred_comparison}. Both models display similar spatial patterns.

\clearpage
\section{Discussion}
\label{sec:discussion}

In this study, we conducted a direct comparison between a Bayesian spatio-temporal modelling framework and a range of statistical or machine learning approaches for predicting norovirus concentrations in wastewater across England. Motivated by the need for improved surveillance tools capable of capturing community-wide transmission dynamics \citep{Walker2024PilotingEngland, Li2023APandemic}, our analysis evaluated predictive performance, uncertainty quantification, and computational cost across the approaches under a consistent spatial cross-validation framework \citep{Pohjankukka2017EstimatingValidation, Ploton2020SpatialModels, Valavi2019BlockCV:Models}.

The strong predictive performance of ensemble tree-based methods, is consistent with their known ability to capture non-linear covariate relationships and complex interaction structures \citep{Hengl2018RandomVariables}. This aligns with previous applications in wastewater surveillance and outbreak prediction \citep{Koureas2021WastewaterMunicipalities, Cho2024ImprovingMethods, Mills2024ThePrevalence}.

However, predictive accuracy alone does not fully determine the suitability of a modelling framework for public-health surveillance. A key contribution of this study is the explicit comparison of uncertainty quantification across modelling approaches. The INLA-SPDE framework \citep{Rue2009ApproximateApproximations, Lindgren2011AnApproach, Blangiardo2015SpatialR-INLA} and the Bayesian GAM \cite{Burkner2017Brms:Stan}, naturally embed probabilistic predictions. 

This property is especially important in public-health decision-making contexts, where underestimation of uncertainty can lead to overconfident predictions and potentially misinformed resource allocation \citep{Zhu2023Spatio-temporalStudy}. The INLA-SPDE framework in particular has been used to identify spatial hotspots and quantify environmental drivers of intestinal infectious diseases at the national level \citep{Zhu2023Spatio-temporalStudy}. In contrast, uncertainty quantification in many machine learning models requires additional procedures such as bootstrap resampling or empirical aggregation across trees \citep{Cho2024ImprovingMethods}.
Random Forest achieved near-nominal empirical coverage and the best 
interval score across all models, suggesting that the prediction intervals were well calibrated. Prediction intervals were constructed using the Quantile Regression Forest framework \citep{Meinshausen2006QuantileForests}, which estimates the full conditional distribution of the response across trees. We note that these intervals do not constitute a calibrated posterior in the Bayesian sense, and should be interpreted as empirical rather than probabilistic \citep{Ennab2022DesigningHealthcare}. The Lasso exhibited particularly low interval coverage, which may have arisen as from a limitation of their frequentist inference where  parameters are treated as fixed at their point estimates rather than marginalised over. The predictive intervals are derived from asymptotic theory rather than full predictive distributions, leading to underestimation of the uncertainty\citep{Wood2017GeneralizedEdition}. 

Computational cost is a further key consideration for real-time operational surveillance, where models must be re-fit frequently as new wastewater data become available \citep{Walker2024PilotingEngland}. Random Forest, XGBoost and GAM completed the full ten-fold cross-validation considerably faster than either Bayesian approach. The INLA-SPDE model was the most computationally demanding,  exceeding the Bayesian GAM  despite both approaches performing  Bayesian inference. This reflects the additional computational burden of the SPDE mesh construction and spatio-temporal field approximation within each fold, rather than Bayesian inference \emph{per se}. The Bayesian GAM, while faster than INLA-SPDE, was still notably slower than the ensemble methods due to MCMC sampling. For short-term forecasting and rapid re-fitting, ensemble ML methods are operationally attractive, whereas Bayesian geostatistical models are perhaps better suited to scheduled analyses where uncertainty quantification and spatial inference are the priority.

The INLA-SPDE model required careful specification of the mesh configuration, prior distributions, and hierarchical structure, which may pose a barrier for practitioners without specialist expertise in spatial statistics \citep{Lindgren2011AnApproach, Blangiardo2015SpatialR-INLA}. Once configured, however, INLA supported continuous-domain prediction at all 33{,}755 LSOAs in England, a capability of clear value for fine-scale public health mapping \citep{Li2023APandemic}. The R package \texttt{inlabru}, which serves as a wrapper for R-INLA, offers a more user-friendly implementation of the INLA-SPDE framework and would perhaps lower the barrier to adoption for applied researchers seeking to implement the INLA-SPDE framework \citep{Bachl2019inlabru}.

Machine learning approaches are generally easier to implement as they require fewer distributional assumptions \citep{Breiman2001RandomForests, Chen2016XGBoost:System}. However, they often do not explicitly account for spatio-temporal structures or hierarchical dependence and rely on covariate proxies (such as coordinates or spatio-temporally varying predictors) to capture spatio-temporal variation \citep{Cabrera2019ModellingModels, Hengl2018RandomVariables}. Exceptions include the Random Forests generalized least squares (RF-GLS) framework and the Bayesian Additive Regression Tree Spatial INLA Modeling and Prediction (BARTSIMP). The RF-GLS can be implemented in the \texttt{RandomForestsGLS} R package \citep{Saha2023RandomData}, which explicitly models spatial dependencies through a Nearest Neighbour Gaussian Process (NNGP) covariance structure directly into the GLS-based splitting criterion, providing a more principled approach rather than including spatial coordinates as covariates. The BARTSIMP framework first fits a Bayesian additive regression tree model to capture covariate effects before modelling the residuals with a Mat\'ern field which is implemented in the INLA-SPDE framework \citep{Jiang2025BARTSIMP}. However, these approaches are limited to modelling spatial dependence and do not support joint spatio-temporal covariance structures. As our data is comprised of repeated wastewater measurements across the same catchment sites over multiple weeks, the repeated coordinate structure miss-specifies the required covariance matrix for these frameworks.

Several limitations should be considered when interpreting these results. First, although our analysis used 3{,}232 samples, the EMHP fortnightly sampling schedule means that the underlying time series at each STW is relatively short, with on average around 21 observations per site over the 45-week study period \citep{Wade2022UnderstandingProgrammes}. This sampling frequency limits the temporal resolution available and may reduce the sensitivity with which short-lived surges viral concentration can be detected. This is an important consideration for norovirus which has a short serial interval of 3 days \citep{Gotz2001ClinicalSweden}, therefore data with higher temporal resolution would allow the models to better capture rapid transmission dynamics. Second, the study covered less than one year of observations, which limits the ability to capture longer-term seasonal dynamics of norovirus transmission \citep{Lopman2009HostWales, Bruggink2010TheRainfall}. Third, although spatial cross-validation was used to mitigate over-optimistic performance estimates \citep{Pohjankukka2017EstimatingValidation, Ploton2020SpatialModels}, residual 
heterogeneity related to sample-handling protocols, sewer-network structure, or local industrial inputs may not be fully captured by the chosen covariates 
\citep{Sims2020FuturePerspectives}. Although the spatial and spatio-temporal random effects in the INLA-SPDE model are designed to absorb such residual catchment-level heterogeneity, the comparator models have no equivalent mechanism to explicitly model spatio-temporal dependencies, which may explain differences observed in the uncertainty calibration across approaches.

Despite these limitations, this study provides one of the first explicit comparisons between Bayesian spatio-temporal modelling and several statistical and machine learning approaches in the context of wastewater-based epidemiology for norovirus surveillance. By leveraging national-scale monitoring data from 152 sewage treatment works, we quantify the trade-offs between predictive performance, uncertainty quantification, and computational cost across a representative set of modelling frameworks.

These findings have implications for the future development of wastewater-based infectious disease surveillance systems. As countries increasingly adopt environmental monitoring to complement clinical surveillance \citep{Walker2024PilotingEngland, Hellmer2014DetectionOutbreaks}, there is a growing need for robust statistical frameworks capable of integrating heterogeneous data sources and providing actionable predictions. Our results suggest that machine learning models may be well-suited for rapid prediction, while Bayesian geostatistical models remain essential when uncertainty quantification and spatial inference at unobserved locations are the objective \citep{Li2023APandemic, Zhu2023Spatio-temporalStudy}. Future work should explore hybrid strategies that combine the predictive flexibility of machine learning with principled uncertainty quantification from Bayesian geostatistical frameworks. Extending such approaches to joint spatio-temporal dependence structures, and evaluating their performance across multiple pathogens and surveillance networks, would further strengthen the evidence base for operational wastewater-based epidemiology.

The frameworks presented here may also be relevant to the surveillance of other priority pathogens. The World Health Organisation has recently updated its list of emerging pathogens to increase awareness and preparedness for future pandemics \citep{Ukoaka2024UpdatedPreparedness}, many of which are emerging or re-emerging in regions where public health infrastructure is less established. Validated modelling tools that have demonstrated effectiveness for one disease can be adapted to support timely detection and intervention for others. Extending these approaches with genomic surveillance and pathogen variant information, could further support the use of wastewater monitoring as an early warning system \citep{Hellmer2014DetectionOutbreaks}.

In summary, this study demonstrates that both Bayesian spatio-temporal models and machine learning algorithms offer valuable tools for modelling norovirus concentrations in wastewater. While ensemble machine learning methods achieved the strongest predictive accuracy, Bayesian geostatistical approaches provided more coherent uncertainty quantification and parameter interpretability, with computational cost varying widely across approaches. These complementary strengths suggest that integrated modelling frameworks may represent the most promising direction for future wastewater-based epidemiology and for strengthening population-level infectious disease surveillance.

\newpage




\clearpage 





\printcredits

\section{Declaration of competing interest}
The authors declare that they have no known competing financial interests or personal relationships that could have appeared to influence the work reported in this paper.

\section{Data availability statement}
Data and code used in this research are openly available and can be found at this \href{https://github.com/erwhite19/Statistical-Learning-Methods-for-Wastewater-Pathogen-Prediction}{github repository.}
\bibliographystyle{cas-model2-names}

\bibliography{references}

@Manual{Simpson2025gratia,
  author  = {Gavin L. Simpson},
  title   = {gratia: Graceful {ggplot}-Based Graphics and Other Functions 
             for {GAM}s Fitted Using {mgcv}},
  year    = {2025},
  url     = {https://gavinsimpson.github.io/gratia/}
}

@article{Lai2013AAge,
    title = {{A norovirus outbreak in a nursing home: Norovirus shedding time associated with age}},
    year = {2013},
    journal = {Journal of Clinical Virology},
    author = {Lai, Chao Chih and Wang, Ying Hsueh and Wu, Ching Yi and Hung, Ching Hsiang and Jiang, Donald Dah Shyong and Wu, Fang Tzy},
    number = {2},
    month = {2},
    pages = {96--101},
    volume = {56},
    publisher = {Elsevier},
    url = {https://www.sciencedirect.com/science/article/pii/S1386653212003927},
    doi = {10.1016/J.JCV.2012.10.011},
    issn = {1386-6532},
    pmid = {23153821}
}

@article{Li2023APandemic,
    title = {{A spatio-temporal framework for modelling wastewater concentration during the COVID-19 pandemic}},
    year = {2023},
    journal = {Environment International},
    author = {Li, Guangquan and Denise, Hubert and Diggle, Peter and Grimsley, Jasmine and Holmes, Chris and James, Daniel and Jersakova, Radka and Mole, Callum and Nicholson, George and Smith, Camila Rangel and Richardson, Sylvia and Rowe, William and Rowlingson, Barry and Torabi, Fatemeh and Wade, Matthew J. and Blangiardo, Marta},
    month = {2},
    volume = {172},
    publisher = {Elsevier Ltd},
    doi = {10.1016/j.envint.2023.107765},
    issn = {18736750},
    pmid = {36709674}
}

@article{Lindgren2011AnApproach,
    title = {{An explicit link between gaussian fields and gaussian markov random fields: The stochastic partial differential equation approach}},
    year = {2011},
    journal = {Journal of the Royal Statistical Society. Series B: Statistical Methodology},
    author = {Lindgren, Finn and Rue, Håvard and Lindstr{\"{o}}m, Johan},
    number = {4},
    month = {9},
    pages = {423--498},
    volume = {73},
    publisher = {John Wiley {\&} Sons, Ltd},
    url = {https://rss.onlinelibrary.wiley.com/doi/10.1111/j.1467-9868.2011.00777.x},
    doi = {10.1111/J.1467-9868.2011.00777.X;PAGE:STRING:ARTICLE/CHAPTER},
    issn = {13697412}
}

@techreport{Rue2009ApproximateApproximations,
    title = {{Approximate Bayesian inference for latent Gaussian models by using integrated nested Laplace approximations}},
    year = {2009},
    booktitle = {J. R. Statist. Soc. B},
    author = {Rue, Håvard and Martino, Sara and Chopin, Nicolas},
    number = {2},
    pages = {319--392},
    volume = {71},
    url = {https://academic.oup.com/jrsssb/article/71/2/319/7092907}
}

@article{Valavi2019BlockCV:Models,
    title = {{blockCV: An r package for generating spatially or environmentally separated folds for k-fold cross-validation of species distribution models}},
    year = {2019},
    journal = {Methods in Ecology and Evolution},
    author = {Valavi, Roozbeh and Elith, Jane and Lahoz-Monfort, José J. and Guillera-Arroita, Gurutzeta},
    number = {2},
    month = {2},
    pages = {225--232},
    volume = {10},
    publisher = {British Ecological Society},
    url = {https://onlinelibrary.wiley.com/doi/abs/10.1111/2041-210X.13107},
    doi = {10.1111/2041-210X.13107;REQUESTEDJOURNAL:JOURNAL:2041210X;WEBSITE:WEBSITE:BESJOURNALS;WGROUP:STRING:PUBLICATION},
    issn = {2041210X}
}

@article{Burkner2017Brms:Stan,
    title = {{brms: An R package for Bayesian multilevel models using Stan}},
    year = {2017},
    journal = {Journal of Statistical Software},
    author = {B{\"{u}}rkner, Paul Christian},
    volume = {80},
    publisher = {American Statistical Association},
    doi = {10.18637/jss.v080.i01},
    issn = {15487660}
}

@misc{CareData.gov.uk,
    title = {{Care home- data.gov.uk}},
    author={Gov UK},
    url = {https://www.data.gov.uk/dataset/7cea3115-1130-41f5-a7b6-3005153acfce/care-homes}
}

@misc{CefasView,
    title = {{Cefas Data Portal - View}},
    author= {Centre for Environment, Fisheries and Aquaculture Science},
    url = {https://data.cefas.co.uk/view/21753}
}

@misc{CensusStatistics,
    title = {{Census - Office for National Statistics}},
    author= {{Office for National Statistics}},
    url = {https://www.ons.gov.uk/census}
}

@article{Ondrikova2023ComparisonSurveillance,
    title = {{Comparison of statistical approaches to predicting norovirus laboratory reports before and during COVID-19: insights to inform public health surveillance}},
    year = {2023},
    journal = {Scientific Reports},
    author = {Ondrikova, Nikola and Clough, Helen and Douglas, Amy and Vivancos, Roberto and Itturiza-Gomara, Miren and Cunliffe, Nigel and Harris, John P.},
    number = {1},
    month = {12},
    volume = {13},
    publisher = {Nature Research},
    doi = {10.1038/s41598-023-48069-6},
    issn = {20452322},
    pmid = {38052922}
}

@article{Arthur2017ContactEcology,
    title = {{Contact structure, mobility, environmental impact and behaviour: the importance of social forces to infectious disease dynamics and disease ecology}},
    year = {2017},
    journal = {Philosophical Transactions of the Royal Society B: Biological Sciences},
    author = {Arthur, Ronan F. and Gurley, Emily S. and Salje, Henrik and Bloomfield, Laura S.P. and Jones, James H.},
    number = {1719},
    month = {3},
    volume = {372},
    publisher = {The Royal Society},
    url = {/doi/pdf/10.1098/rstb.2016.0454},
    doi = {10.1098/RSTB.2016.0454},
    issn = {14712970},
    pmid = {28289265}
}

@misc{CORINEService,
    title = {{CORINE Land Cover — Copernicus Land Monitoring Service}},
    author= {{Copernicus Land Monitoring Service}},
    url = {https://land.copernicus.eu/en/products/corine-land-cover}
}

@misc{COVID-19Reports,
    title = {{COVID-19 Community Mobility Reports}},
    author= {Google},
    url = {https://www.google.com/covid19/mobility/}
}

@article{Ennab2022DesigningHealthcare,
    title = {{Designing an Interpretability-Based Model to Explain the Artificial Intelligence Algorithms in Healthcare}},
    year = {2022},
    journal = {Diagnostics},
    author = {Ennab, Mohammad and McHeick, Hamid},
    number = {7},
    month = {7},
    pages = {1557},
    volume = {12},
    publisher = {Multidisciplinary Digital Publishing Institute (MDPI)},
    url = {https://pmc.ncbi.nlm.nih.gov/articles/PMC9319389/},
    doi = {10.3390/DIAGNOSTICS12071557},
    issn = {20754418},
    pmid = {35885463}
}

@article{Hellmer2014DetectionOutbreaks,
    title = {{Detection of pathogenic viruses in sewage provided early warnings of hepatitis A virus and norovirus outbreaks}},
    year = {2014},
    journal = {Applied and Environmental Microbiology},
    author = {Hellm{\'{e}}r, Maria and Pax{\'{e}}us, Nicklas and Magnius, Lars and Enache, Lucica and Arnholm, Birgitta and Johansson, Annette and Bergstr{\"{o}}m, Tomas and Norder, Heléne},
    number = {21},
    pages = {6771--6781},
    volume = {80},
    publisher = {American Society for Microbiology},
    url = {https://pubmed.ncbi.nlm.nih.gov/25172863/},
    doi = {10.1128/AEM.01981-14,},
    issn = {10985336},
    pmid = {25172863}
}

@misc{DownloadsGOV.UK,
    title = {{Downloads - GOV.UK}},
     author= {GOV.UK},
    url = {https://get-information-schools.service.gov.uk/Downloads}
}

@article{Simmons2013DurationCDC,
    title = {{Duration of Immunity to Norovirus Gastroenteritis - Volume 19, Number 8—August 2013 - Emerging Infectious Diseases journal - CDC}},
    year = {2013},
    journal = {Emerging Infectious Diseases},
    author = {Simmons, Kirsten and Gambhir, Manoj and Leon, Juan and Lopman, Ben},
    number = {8},
    month = {8},
    pages = {1260--1267},
    volume = {19},
    url = {https://wwwnc.cdc.gov/eid/article/19/8/13-0472_article},
    doi = {10.3201/EID1908.130472},
    issn = {10806040},
    pmid = {23876612}
}

@article{Tam2016EconomicKingdom,
    title = {{Economic Cost of Campylobacter, Norovirus and Rotavirus Disease in the United Kingdom}},
    year = {2016},
    journal = {PLOS ONE},
    author = {Tam, Clarence C. and O'Brien, Sarah J.},
    number = {2},
    month = {2},
    pages = {e0138526},
    volume = {11},
    publisher = {Public Library of Science},
    url = {https://journals.plos.org/plosone/article?id=10.1371/journal.pone.0138526},
    doi = {10.1371/JOURNAL.PONE.0138526},
    issn = {1932-6203},
    pmid = {26828435}
}

@article{Lian2019Epidemiology20142017,
    title = {{Epidemiology of Norovirus Outbreaks Reported to the Public Health Emergency Event Surveillance System, China, 2014–2017}},
    year = {2019},
    journal = {Viruses},
    author = {Lian, Yiyao and Wu, Shuyu and Luo, Li and Lv, Bin and Liao, Qiaohong and Li, Zhongjie and Rainey, Jeanette J. and Hall, Aron J. and Ran, Lu},
    number = {4},
    month = {4},
    pages = {342},
    volume = {11},
    publisher = {MDPI AG},
    url = {https://pmc.ncbi.nlm.nih.gov/articles/PMC6520956/},
    doi = {10.3390/V11040342},
    issn = {19994915},
    pmid = {30979014}
}

@misc{ERA5Present,
    title = {{ERA5 hourly data on single levels from 1940 to present}},
    author= {{Climate Data Store}},
    url = {https://cds.climate.copernicus.eu/datasets/reanalysis-era5-single-levels?tab=overview}
}

@article{Pohjankukka2017EstimatingValidation,
    title = {{Estimating the prediction performance of spatial models via spatial k-fold cross validation}},
    year = {2017},
    journal = {International Journal of Geographical Information Science},
    author = {Pohjankukka, Jonne and Pahikkala, Tapio and Nevalainen, Paavo and Heikkonen, Jukka},
    number = {10},
    month = {10},
    pages = {2001--2019},
    volume = {31},
    publisher = {Taylor {\&} Francis},
    url = {https://www.tandfonline.com/doi/pdf/10.1080/13658816.2017.1346255},
    doi = {10.1080/13658816.2017.1346255},
    issn = {13623087},
    arxivId = {2005.14263}
}

@article{Wood2017GeneralizedEdition,
    title = {{Generalized additive models: An introduction with R, second edition}},
    year = {2017},
    journal = {Generalized Additive Models: An Introduction with R, Second Edition},
    author = {Wood, Simon N.},
    month = {1},
    pages = {1--476},
    publisher = {CRC Press},
    isbn = {9781498728348},
    doi = {10.1201/9781315370279}
}

@misc{Friedman2010Glmnet:Models,
    title = {{glmnet: Lasso and elastic-net regularized generalized linear models}},
    year = {2010},
    booktitle = {Journal of Statistical Software},
    author = {Friedman, Jerome and Hastie, Trevor and Tibshirani, Rob},
    number = {1},
    volume = {33},
    doi = {10.32614/CRAN.package.glmnet},
    issn = {15487660}
}

@article{Wang2023GlobalMeta-analysis,
    title = {{Global prevalence of asymptomatic norovirus infection in outbreaks: a systematic review and meta-analysis}},
    year = {2023},
    journal = {BMC Infectious Diseases},
    author = {Wang, Jun and Gao, Zhao and Yang, Zu rong and Liu, Kun and Zhang, Hui},
    number = {1},
    month = {12},
    volume = {23},
    publisher = {BioMed Central Ltd},
    url = {https://pubmed.ncbi.nlm.nih.gov/37700223/},
    doi = {10.1186/S12879-023-08519-Y,},
    issn = {14712334},
    pmid = {37700223}
}

@article{Qi2018GlobalMeta-analysis,
    title = {{Global Prevalence of Asymptomatic Norovirus Infection: A Meta-analysis}},
    year = {2018},
    journal = {EClinicalMedicine},
    author = {Qi, Rui and Huang, Yu ting and Liu, Jian wei and Sun, Yue and Sun, Xi feng and Han, Hui Ju and Qin, Xiang Rong and Zhao, Min and Wang, Li jun and Li, Wenqian and Li, Jun hong and Chen, Cong and Yu, Xue Jie},
    month = {8},
    pages = {50--58},
    volume = {2-3},
    publisher = {Lancet Publishing Group},
    url = {https://pubmed.ncbi.nlm.nih.gov/31193628/},
    doi = {10.1016/j.eclinm.2018.09.001},
    issn = {25895370}
}

@article{Lopman2009HostWales,
    title = {{Host, weather and virological factors drive norovirus epidemiology: Time-series analysis of laboratory surveillance data in England and Wales}},
    year = {2009},
    journal = {PLoS ONE},
    author = {Lopman, Ben and Armstrong, Ben and Atchison, Christina and Gray, Jim J.},
    number = {8},
    month = {8},
    volume = {4},
    doi = {10.1371/journal.pone.0006671},
    issn = {19326203},
    pmid = {19701458}
}

@article{Cho2024ImprovingMethods,
    title = {{Improving Norovirus Outbreak Prediction through Feature Selection using Machine Learning Methods}},
    year = {2024},
    journal = {J. Korean Soc. Ind. Appl. Math},
    author = {Cho, Giphil and Seo, Jeonghwa and Lee, Hyojung},
    number = {4},
    pages = {210--225},
    volume = {28},
    url = {https://doi.org/10.12941/jksiam.2024.28.210},
    doi = {10.12941/jksiam.2024.28.210}
}

@misc{Lockdown_2021_2022Storage,
    title = {COVID-19 Timeline},
    author= {Keoghs},
    url = {https://keoghs.co.uk/keoghs-insight/covid-19-timeline}
}

@article{Wood2001Mgcv:R,
    title = {{mgcv: GAMs and generalized ridge regression for R}},
    year = {2001},
    journal = {R News},
    author = {Wood, S N},
    volume = {1}
}

@article{Cabrera2019ModellingModels,
    title = {{Modelling spatio-temporal data of dengue fever using generalized additive mixed models}},
    year = {2019},
    journal = {Spatial and Spatio-temporal Epidemiology},
    author = {Cabrera, M. and Taylor, G.},
    month = {2},
    pages = {1--13},
    volume = {28},
    publisher = {Elsevier},
    url = {https://www.sciencedirect.com/science/article/pii/S1877584517301363},
    doi = {10.1016/J.SSTE.2018.11.006},
    issn = {1877-5845},
    pmid = {30739650}
}

@misc{NationalGOV.UK,
    title = {{National norovirus and rotavirus report, week 31 report: data to week 30 (data up to 27 July 2025) - GOV.UK}},
    author= {UKHSA},
    url = {https://www.gov.uk/government/statistics/national-norovirus-and-rotavirus-surveillance-reports-2025-to-2026-season/national-norovirus-and-rotavirus-report-week-31-report-data-to-week-30-data-up-to-27-july-2025#background-information}
}

@article{Gholipour2022OccurrenceReview,
    title = {{Occurrence of viruses in sewage sludge: A systematic review}},
    year = {2022},
    journal = {Science of The Total Environment},
    author = {Gholipour, Sahar and Ghalhari, Mohammad Rezvani and Nikaeen, Mahnaz and Rabbani, Davarkhah and Pakzad, Parichehr and Miranzadeh, Mohammad Bagher},
    month = {6},
    pages = {153886},
    volume = {824},
    publisher = {Elsevier},
    url = {https://www.sciencedirect.com/science/article/pii/S0048969722009780?via%3Dihub},
    doi = {10.1016/J.SCITOTENV.2022.153886},
    issn = {0048-9697},
    pmid = {35182626}
}

@article{Simpson2017PenalisingPriors,
    title = {{Penalising Model Component Complexity: A Principled, Practical Approach to Constructing Priors}},
    year = {2017},
    journal = {https://doi.org/10.1214/16-STS576},
    author = {Simpson, Daniel and Rue, Håvard and Riebler, Andrea and Martins, Thiago G. and S{\o}rbye, Sigrunn H.},
    number = {1},
    month = {2},
    pages = {1--28},
    volume = {32},
    publisher = {Institute of Mathematical Statistics},
    url = {https://projecteuclid.org/journals/statistical-science/volume-32/issue-1/Penalising-Model-Component-Complexity--A-Principled-Practical-Approach-to/10.1214/16-STS576.full},
    doi = {10.1214/16-STS576},
    issn = {0883-4237},
    arxivId = {1403.4630}
}

@article{Walker2024PilotingEngland,
    title = {{Piloting wastewater-based surveillance of norovirus in England}},
    year = {2024},
    journal = {Water Research},
    author = {Walker, David I. and Witt, Jessica and Rostant, Wayne and Burton, Robert and Davison, Vicki and Ditchburn, Jackie and Evens, Nicholas and Godwin, Reg and Heywood, Jane and Lowther, James A. and Peters, Nancy and Porter, Jonathan and Posen, Paulette and Wickens, Tyler and Wade, Matthew J.},
    month = {10},
    volume = {263},
    publisher = {Elsevier Ltd},
    doi = {10.1016/j.watres.2024.122152},
    issn = {18792448},
    pmid = {39096810}
}

@article{Hengl2018RandomVariables,
    title = {{Random forest as a generic framework for predictive modeling of spatial and spatio-temporal variables}},
    year = {2018},
    journal = {PeerJ},
    author = {Hengl, Tomislav and Nussbaum, Madlene and Wright, Marvin N. and Heuvelink, Gerard B.M. and Gr{\"{a}}ler, Benedikt},
    number = {8},
    month = {8},
    pages = {e5518},
    volume = {2018},
    publisher = {PeerJ Inc.},
    url = {https://peerj.com/articles/5518},
    doi = {10.7717/PEERJ.5518/SUPP-1},
    issn = {21678359},
    pmid = {30186691}
}

@article{Harris2017Re-assessingPopulation,
    title = {{Re-assessing the total burden of norovirus circulating in the United Kingdom population}},
    year = {2017},
    journal = {Vaccine},
    author = {Harris, John P. and Iturriza-Gomara, Miren and O'Brien, Sarah J.},
    number = {6},
    month = {2},
    pages = {853--855},
    volume = {35},
    publisher = {Elsevier},
    url = {https://www.sciencedirect.com/science/article/pii/S0264410X17300142?via%3Dihub},
    doi = {10.1016/J.VACCINE.2017.01.009},
    issn = {0264-410X},
    pmid = {28094075}
}

@book{Blangiardo2015SpatialR-INLA,
    title = {{Spatial and spatio-temporal Bayesian models with R-INLA}},
    year = {2015},
    author = {Blangiardo, Marta. and Cameletti, Michela.},
    publisher = {John Wiley and Sons, Inc.},
    isbn = {9781118326558}
}

@article{Ploton2020SpatialModels,
    title = {{Spatial validation reveals poor predictive performance of large-scale ecological mapping models}},
    year = {2020},
    journal = {Nature Communications 2020 11:1},
    author = {Ploton, Pierre and Mortier, Frédéric and R{\'{e}}jou-M{\'{e}}chain, Maxime and Barbier, Nicolas and Picard, Nicolas and Rossi, Vivien and Dormann, Carsten and Cornu, Guillaume and Viennois, Gaëlle and Bayol, Nicolas and Lyapustin, Alexei and Gourlet-Fleury, Sylvie and P{\'{e}}lissier, Raphaël},
    number = {1},
    month = {9},
    pages = {1--11},
    volume = {11},
    publisher = {Nature Publishing Group},
    url = {https://www.nature.com/articles/s41467-020-18321-y},
    doi = {10.1038/s41467-020-18321-y},
    issn = {2041-1723},
    pmid = {32917875}
}

@article{Zhu2023Spatio-temporalStudy,
    title = {{Spatio-temporal pattern and associate factors of intestinal infectious diseases in Zhejiang Province, China, 2008–2021: a Bayesian modeling study}},
    year = {2023},
    journal = {BMC Public Health},
    author = {Zhu, Zhixin and Feng, Yan and Gu, Lanfang and Guan, Xifei and Liu, Nawen and Zhu, Xiaoxia and Gu, Hua and Cai, Jian and Li, Xiuyang},
    number = {1},
    month = {12},
    pages = {1--12},
    volume = {23},
    publisher = {BioMed Central Ltd},
    url = {https://link.springer.com/articles/10.1186/s12889-023-16552-4},
    doi = {10.1186/S12889-023-16552-4/FIGURES/5},
    issn = {14712458},
    pmid = {37644452}
}

@misc{Patel2008SystematicGastroenteritis,
    title = {{Systematic literature review of role of noroviruses in sporadic gastroenteritis}},
    year = {2008},
    booktitle = {Emerging Infectious Diseases},
    author = {Patel, Manish M. and Widdowson, Marc Alain and Glass, Roger I. and Akazawa, Kenichiro and Vinj{\'{e}}, Jan and Parashar, Umesh D.},
    number = {8},
    month = {8},
    pages = {1224--1231},
    volume = {14},
    doi = {10.3201/eid1408.071114},
    issn = {10806040},
    pmid = {18680645}
}

@article{Bruggink2010TheRainfall,
    title = {{The incidence of norovirus-associated gastroenteritis outbreaks in victoria, Australia (2002-2007) and their relationship with rainfall}},
    year = {2010},
    journal = {International Journal of Environmental Research and Public Health},
    author = {Bruggink, Leesa D. and Marshall, John A.},
    number = {7},
    pages = {2822--2827},
    volume = {7},
    publisher = {MDPI},
    doi = {10.3390/ijerph7072822},
    issn = {16604601},
    pmid = {20717541}
}

@article{Mills2024ThePrevalence,
    title = {{The utility of wastewater surveillance for monitoring SARS-CoV-2 prevalence}},
    year = {2024},
    journal = {PNAS Nexus},
    author = {Mills, Cathal and Chadeau-Hyam, Marc and Elliott, Paul and Donnelly, Christl A.},
    number = {10},
    month = {10},
    volume = {3},
    publisher = {National Academy of Sciences},
    doi = {10.1093/pnasnexus/pgae438},
    issn = {27526542},
    pmid = {39474505}
}

@misc{TimelineGovernment,
    title = {{Timeline of UK government coronavirus lockdowns and restrictions | Institute for Government}},
    author= {{Institute for Government}},
    url = {https://www.instituteforgovernment.org.uk/data-visualisation/timeline-coronavirus-lockdowns}
}

@article{Tsang2018TransmissibilityChina,
    title = {{Transmissibility of Norovirus in Urban versus Rural Households in a Large Community Outbreak in China}},
    year = {2018},
    journal = {Epidemiology (Cambridge, Mass.)},
    author = {Tsang, Tim K. and Chen, Tian Mu and Longini, Ira M. and Halloran, M. Elizabeth and Wu, Ying and Yang, Yang},
    number = {5},
    pages = {675},
    volume = {29},
    publisher = {Lippincott Williams and Wilkins},
    url = {https://pmc.ncbi.nlm.nih.gov/articles/PMC6066449/},
    doi = {10.1097/EDE.0000000000000855},
    issn = {15315487},
    pmid = {29847497}
}

@article{Wade2022UnderstandingProgrammes,
    title = {{Understanding and managing uncertainty and variability for wastewater monitoring beyond the pandemic: Lessons learned from the United Kingdom national COVID-19 surveillance programmes}},
    year = {2022},
    journal = {Journal of Hazardous Materials},
    author = {Wade, Matthew J. and Lo Jacomo, Anna and Armenise, Elena and Brown, Mathew R. and Bunce, Joshua T. and Cameron, Graeme J. and Fang, Zhou and Farkas, Kata and Gilpin, Deidre F. and Graham, David W. and Grimsley, Jasmine M.S. and Hart, Alwyn and Hoffmann, Till and Jackson, Katherine J. and Jones, David L. and Lilley, Chris J. and McGrath, John W. and McKinley, Jennifer M. and McSparron, Cormac and Nejad, Behnam F. and Morvan, Mario and Quintela-Baluja, Marcos and Roberts, Adrian M.I. and Singer, Andrew C. and Souque, Célia and Speight, Vanessa L. and Sweetapple, Chris and Walker, David and Watts, Glenn and Weightman, Andrew and Kasprzyk-Hordern, Barbara},
    month = {2},
    volume = {424},
    publisher = {Elsevier B.V.},
    doi = {10.1016/j.jhazmat.2021.127456},
    issn = {18733336},
    pmid = {34655869}
}

@article{Ukoaka2024UpdatedPreparedness,
    title = {{Updated WHO list of emerging pathogens for a potential future pandemic: Implications for public health and global preparedness}},
    year = {2024},
    journal = {Le Infezioni in Medicina},
    author = {Ukoaka, Bonaventure Michael and Okesanya, Olalekan John and Daniel, Faithful Miebaka and Ahmed, Mohammed Mustaf and Udam, Ntishor Gabriel and Wagwula, Precious Miracle and Adigun, Olaniyi Abideen and Udoh, Raphael Augustine and Peter, Iniubong Godswill and Lawal, Haleema},
    number = {4},
    pages = {463},
    volume = {32},
    publisher = {EDIMES Edizioni Medico Scientifiche},
    url = {https://pmc.ncbi.nlm.nih.gov/articles/PMC11627490/},
    doi = {10.53854/LIIM-3204-5},
    issn = {11249390},
    pmid = {39660154}
}

@article{Koureas2021WastewaterMunicipalities,
    title = {{Wastewater monitoring as a supplementary surveillance tool for capturing SARS-COV-2 community spread. A case study in two Greek municipalities}},
    year = {2021},
    journal = {Environmental Research},
    author = {Koureas, Michalis and Amoutzias, Grigoris D. and Vontas, Alexandros and Kyritsi, Maria and Pinaka, Ourania and Papakonstantinou, Argyrios and Dadouli, Katerina and Hatzinikou, Marina and Koutsolioutsou, Anastasia and Mouchtouri, Varvara A. and Speletas, Matthaios and Tsiodras, Sotirios and Hadjichristodoulou, Christos},
    month = {9},
    pages = {111749},
    volume = {200},
    publisher = {Academic Press},
    doi = {10.1016/J.ENVRES.2021.111749},
    issn = {0013-9351},
    pmid = {34310965}
}

@article{Chen2016XGBoost:System,
    title = {{XGBoost: A Scalable Tree Boosting System}},
    year = {2016},
    journal = {Proceedings of the ACM SIGKDD International Conference on Knowledge Discovery and Data Mining},
    author = {Chen, Tianqi and Guestrin, Carlos},
    month = {3},
    pages = {785--794},
    volume = {13-17-August-2016},
    publisher = {Association for Computing Machinery},
    url = {https://arxiv.org/pdf/1603.02754},
    isbn = {9781450342322},
    doi = {10.1145/2939672.2939785},
    arxivId = {1603.02754}
}

@article{Liaw2002Classification,
    title = {{Classification and Regression by randomForest}},
    year = {2002},
    journal = {R News},
    author = {Liaw, Andy and Wiener, Matthew},
    number = {3},
    pages = {18--22},
    volume = {2},
    url = {https://CRAN.R-project.org/doc/Rnews/}
}

@article{Gelman2008WeaklyAnalyses,
    title = {{A weakly informative default prior distribution for logistic and other regression models}},
    year = {2008},
    journal = {Annals of Applied Statistics},
    author = {Gelman, Andrew and Jakulin, Aleks and Pittau, Maria Grazia and Su, Yu Sung},
    number = {4},
    month = {12},
    pages = {1360--1383},
    volume = {2},
    publisher = {Institute of Mathematical Statistics},
    url = {https://projecteuclid.org/euclid.aoas/1231424214},
    doi = {10.1214/08-AOAS191},
    issn = {19326157},
    arxivId = {0901.4011}
}

@article{Vehtari2021Rank,
    title = {{Rank-Normalization, Folding, and Localization: An Improved Rhat for Assessing Convergence of MCMC (with Discussion)}},
    year = {2021},
    journal = {Bayesian Analysis},
    author = {Vehtari, Aki and Gelman, Andrew and Simpson, Daniel and Carpenter, Bob and B{\"{u}}rkner, Paul Christian},
    number = {2},
    month = {6},
    pages = {667--718},
    volume = {16},
    publisher = {International Society for Bayesian Analysis},
    url = {https://projecteuclid.org/journals/bayesian-analysis/volume-16/issue-2/Rank-Normalization-Folding-and-Localization--An-Improved-R̂-for/10.1214/20-BA1221.full},
    doi = {10.1214/20-BA1221},
    issn = {19360975},
    arxivId = {1903.08008}
}

@article{Breiman2001RandomForests,
    title = {{Random Forests}},
    year = {2001},
    journal = {Machine Learning},
    author = {Breiman, Leo},
    number = {1},
    month = {10},
    pages = {5--32},
    volume = {45},
    publisher = {Springer},
    url = {https://link.springer.com/article/10.1023/A:1010933404324},
    doi = {10.1023/A:1010933404324},
    issn = {08856125}
}

@article{Wood2003ThinSplines,
    title = {{Thin plate regression splines}},
    year = {2003},
    journal = {Journal of the Royal Statistical Society. Series B: Statistical Methodology},
    author = {Wood, Simon N.},
    number = {1},
    month = {2},
    pages = {95--114},
    volume = {65},
    publisher = {John Wiley {\&} Sons, Ltd},
    url = {https://rss.onlinelibrary.wiley.com/doi/10.1111/1467-9868.00374},
    doi = {10.1111/1467-9868.00374},
    issn = {13697412}
}

@article{Saha2023RandomData,
    title = {{Random Forests for Spatially Dependent Data}},
    year = {2023},
    journal = {Journal of the American Statistical Association},
    author = {Saha, Arkajyoti and Basu, Sumanta and Datta, Abhirup},
    number = {541},
    pages = {665--683},
    volume = {118},
    publisher = {Taylor {\&} Francis},
    url = {https://www.tandfonline.com/doi/full/10.1080/01621459.2021.1950003},
    doi = {10.1080/01621459.2021.1950003},
    issn = {1537274X}
}

@article{Gotz2001ClinicalSweden,
    title = {{Clinical spectrum and transmission characteristics of infection with 
    Norwalk-like virus: findings from a large community outbreak in Sweden}},
    year = {2001},
    journal = {Clinical Infectious Diseases},
    author = {G{\"{o}}tz, Hans and Ekdahl, Karl and Lindback, Johan and de Jong, 
    Birgitta and Hedlund, Karl-Olof and Giesecke, Johan},
    number = {5},
    pages = {622--628},
    volume = {33},
    publisher = {Oxford University Press},
    doi = {10.1086/322608},
    issn = {10584838}
}

@article{Sims2020FuturePerspectives,
    title = {{Future perspectives of wastewater-based epidemiology: Monitoring infectious 
    disease spread and resistance to the community level}},
    year = {2020},
    journal = {Environment International},
    author = {Sims, Natalie and Kasprzyk-Hordern, Barbara},
    month = {6},
    pages = {105689},
    volume = {139},
    publisher = {Elsevier},
    doi = {10.1016/j.envint.2020.105689},
    issn = {01604120}
}

@techreport{hoffmann_wastewater_2022,
    type = {preprint},
    title = {Wastewater catchment areas in {Great} {Britain}},
    url = {https://essopenarchive.org/doi/full/10.1002/essoar.10510612.2},
    doi = {10.1002/essoar.10510612.2},
    language = {en},
    urldate = {2023-11-28},
    institution = {Environmental Sciences},
    author = {Hoffmann, Till and Bunney, Sarah and Kasprzyk-Hordern, Barbara and Singer, Andrew},
    month = feb,
    year = {2022},
}

@article{Jiang2025BARTSIMP,
    title = {{BARTSIMP: Flexible Spatial Covariate Modeling and Prediction 
              Using Bayesian Additive Regression Trees}},
    year = {2025},
    journal = {Spatial and Spatio-temporal Epidemiology},
    author = {Jiang, A.Z. and Wakefield, Jon},
    volume = {55},
    pages = {100757},
    publisher = {Elsevier},
    url = {https://www.sciencedirect.com/science/article/pii/S1877584525000486},
    doi = {10.1016/j.sste.2025.100757},
    issn = {1877-5845}
}

@article{Bachl2019inlabru,
    title = {{inlabru: an R package for Bayesian spatial modelling 
              from ecological survey data}},
    year = {2019},
    journal = {Methods in Ecology and Evolution},
    author = {Bachl, Fabian E. and Lindgren, Finn and Borchers, David L. 
              and Illian, Janine B.},
    number = {6},
    month = {6},
    pages = {760--766},
    volume = {10},
    publisher = {British Ecological Society},
    url = {https://doi.org/10.1111/2041-210X.13168},
    doi = {10.1111/2041-210X.13168},
    issn = {2041-210X}
}

@Manual{Hijmans2024terra,
    title = {{terra: Spatial Data Analysis}},
    author = {Hijmans, Robert J.},
    year = {2024},
    note = {R package version 1.7-78},
    url = {https://CRAN.R-project.org/package=terra},
}

@misc{ONS2023LookupTable,
    title = {{Postcode to OA (2021) to LSOA to MSOA to LAD (August 2023) 
              Best Fit Lookup in the UK}},
    year = {2023},
    author = {{Office for National Statistics}},
    publisher = {Office for National Statistics},
    url = {https://geoportal.statistics.gov.uk/datasets/3770c5e8b0c24f1dbe6d2fc6b46a0b18/about},
    note = {Accessed: 2026-07-07}
}

@article{Meinshausen2006QuantileForests,
    title = {{Quantile Regression Forests}},
    year = {2006},
    journal = {Journal of Machine Learning Research},
    author = {Meinshausen, Nicolai and Ridgeway, Greg},
    number = {6},
    pages = {983--999},
    volume = {7},
    url = {https://www.jmlr.org/papers/v7/meinshausen06a.html},
    issn = {1532-4435}
}

@article{Bowes2024WBE,
  title   = {Wastewater-based epidemiology to assess environmentally influenced disease},
  author  = {Bowes, Devin A. and Driver, Erin M. and Choi, Phil M. and Barcelo, Dami{\`a} and Beamer, Paloma I.},
  journal = {Journal of Exposure Science \& Environmental Epidemiology},
  volume  = {34},
  pages   = {387--388},
  year    = {2024},
  doi     = {10.1038/s41370-024-00683-w}
}

@article{Alirol2011Urbanisation,
    title   = {Urbanisation and infectious diseases in a globalised world},
    author  = {Alirol, Emilie and Getaz, Laurent and Stoll, Beat and Chappuis, Fran{\c{c}}ois and Loutan, Louis},
    journal = {The Lancet Infectious Diseases},
    volume  = {11},
    number  = {2},
    pages   = {131--141},
    year    = {2011},
    doi     = {10.1016/S1473-3099(10)70223-1}
}

@article{Fowler2013TheNitrogen,
    title   = {The global nitrogen cycle in the twenty-first century},
    author  = {Fowler, David and Coyle, Mhairi and Skiba, Ute and Sutton, Mark A. and Cape, J. Neil and Reis, Stefan and others},
    journal = {Philosophical Transactions of the Royal Society B},
    volume  = {368},
    number  = {1621},
    pages   = {20130164},
    year    = {2013},
    doi     = {10.1098/rstb.2013.0164}
}

@article{roberts_data_2022,
	title = {Data normalisation of {RT}-{qPCR} data for detection of {SARS}-{CoV}-2 in wastewater},
	url = {https://www.protocols.io/view/data-normalisation-of-rt-qpcr-data-for-detection-o-b4eqqtdw},
	language = {en},
	urldate = {2025-09-15},
	author = {Roberts, Adrian and Fang, Zhou and Mayer, Claus-Dieter and Frantsuzova, Anastasia and Cameron, Graeme J. and Scorza, Livia C. T.},
	month = mar,
	year = {2022},
}

@article{Gneiting2007,
  author    = {Gneiting, Tilmann and Raftery, Adrian E.},
  title     = {Strictly Proper Scoring Rules, Prediction, and Estimation},
  journal   = {Journal of the American Statistical Association},
  year      = {2007},
  volume    = {102},
  number    = {477},
  pages     = {359--378},
  doi       = {10.1198/016214506000001437},
  url       = {https://doi.org/10.1198/016214506000001437}
}



\newpage
\appendix
\section{SPDE Approach}\label{appendix:spde}

The SPDE approach, introduced by \cite{Lindgren2011AnApproach}, provides a 
computationally efficient representation of a continuously indexed Gaussian 
random field by expressing it as a discretely indexed Gaussian Markov Random 
Field (GMRF). Specifically, the spatial field is defined as the solution to 
the SPDE

\begin{equation}
    (\kappa^2 - \Delta)^{\alpha/2} \tau u(\mathbf{s}) = \mathcal{W}(\mathbf{s}),
    \label{eq:spde}
\end{equation}

where $\kappa > 0$ controls the spatial range, $\tau > 0$ controls the 
marginal variance, $\Delta$ is the Laplacian operator, and 
$\mathcal{W}(\mathbf{s})$ is a Gaussian spatial white noise process. The 
solution to \eqref{eq:spde} gives a Matérn covariance structure, as 
specified in Section~\ref{sec:bayesian_model}, with smoothness parameter 
$\lambda_\nu = \alpha - d/2$, where $d$ is the spatial dimension. The 
spatial field is approximated using a finite set of weighted basis functions 
defined at the vertices of a triangulated mesh over the study region, represented by:

\begin{equation}
    u(\mathbf{s}) \approx \sum_{k=1}^{K} \psi_k(\mathbf{s})\, \tilde{u}_k,
\end{equation}

where $\psi_k(\mathbf{s})$ are piecewise linear basis functions and 
$\tilde{u}_k$ are Gaussian distributed weights at the $K$ mesh vertices. 
This representation induces sparsity in the precision matrix of the GMRF, 
enabling efficient Bayesian inference when combined with the INLA 
framework \citep{Rue2009ApproximateApproximations}.

\clearpage
\section{Semivariogram}\label{appendix:variogram}

\noindent
\begin{center}
    \includegraphics[width=0.5\linewidth]{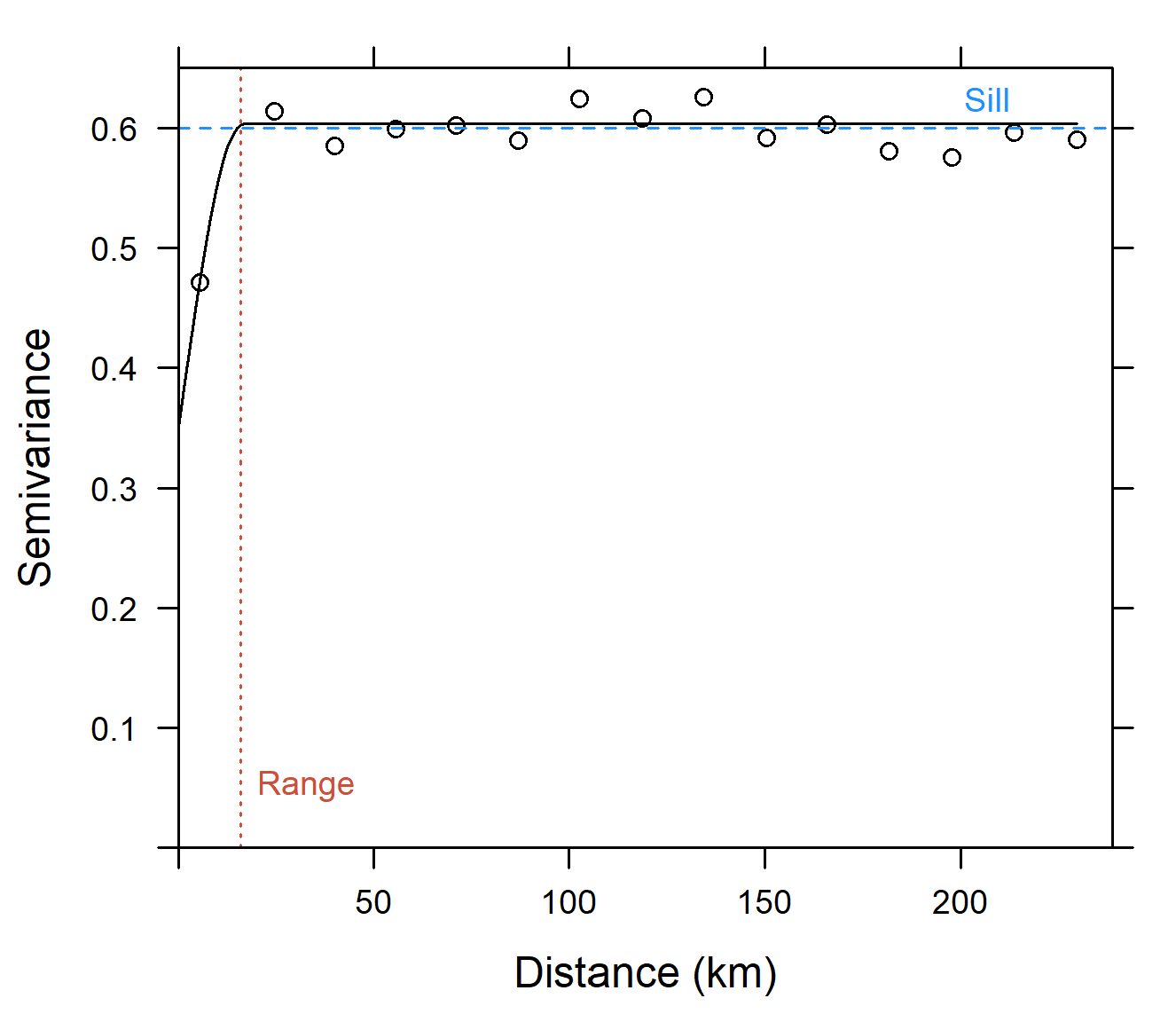}
    \captionof{figure}{Empirical semivariogram of the spatial field.}
    \label{fig:variogram}
\end{center}

\clearpage
\section{Bayesian GAM Convergence Assessment}\label{appendix:trace}

Figure~\ref{fig:fold1}, \ref{fig:fold5}, and \ref{fig:fold10} 
display the posterior distributions and trace plots for a representative selection of folds (Folds 1, 5, and 10) from the 10-fold cross-validation of the Bayesian GAM model. All chains show good mixing and convergence, with trace plots exhibiting stable stationarity and no evidence of divergence. The remaining folds displayed similar convergence diagnostics and are not shown for brevity.

\begin{figure}[H]
    \centering
    \includegraphics[width=\textwidth]{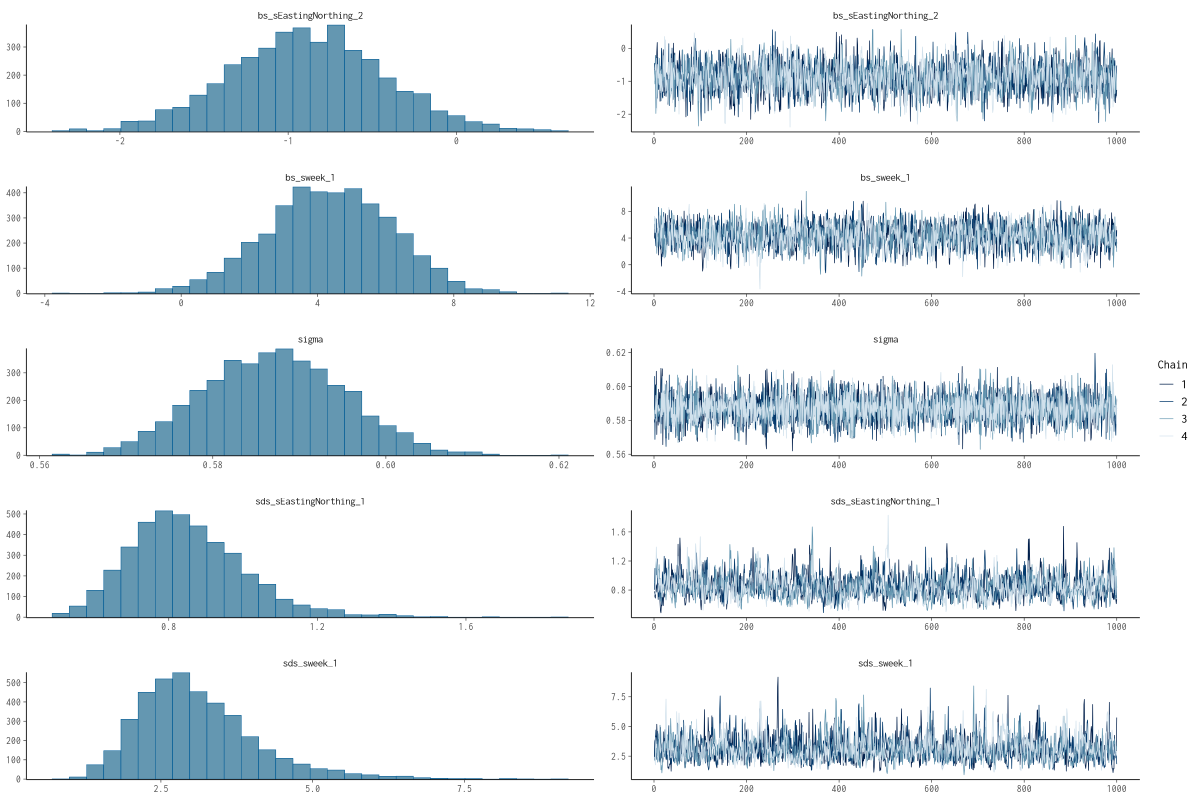}
    \caption{Trace plots and posterior distributions for Fold 1.}
    \label{fig:fold1}
\end{figure}

\begin{figure}[H]
    \centering
    \includegraphics[width=\textwidth]{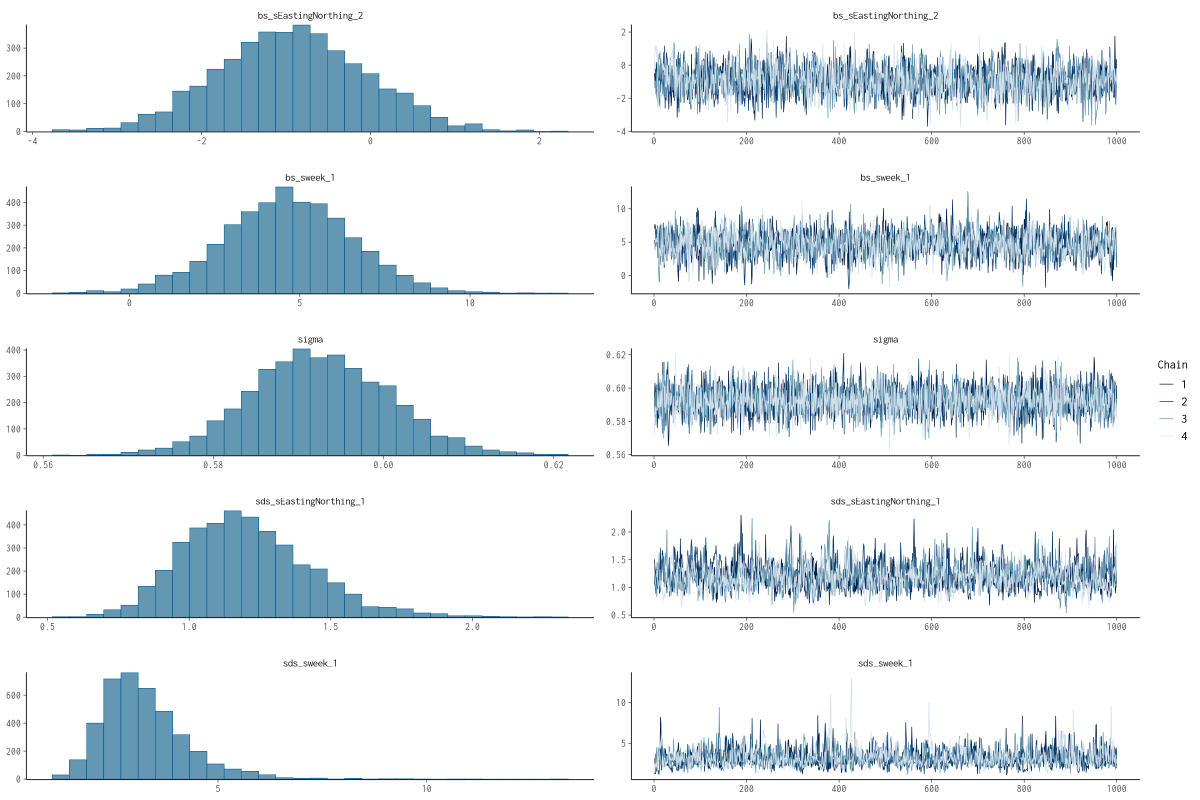}
    \caption{Trace plots and posterior distributions for Fold 5.}
    \label{fig:fold5}
\end{figure}

\begin{figure}[H]
    \centering
    \includegraphics[width=\textwidth]{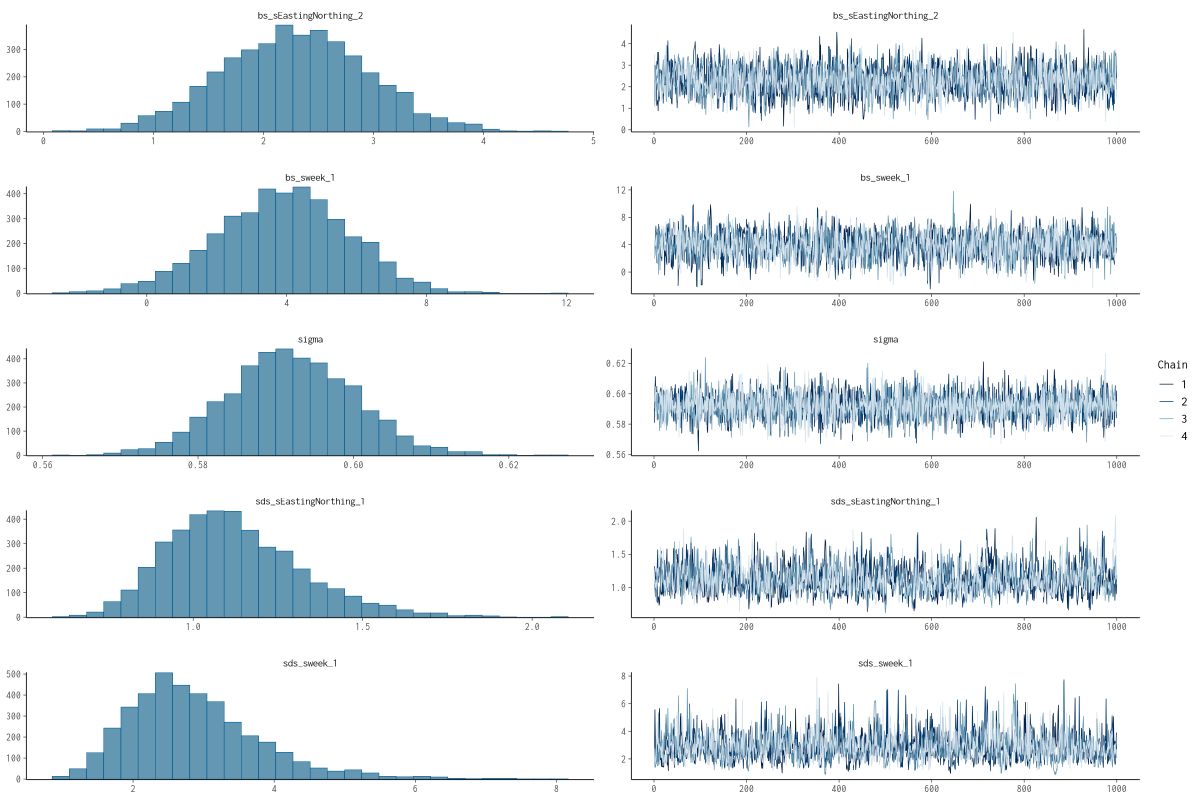}
    \caption{Trace plots and posterior distributions for Fold 10.}
    \label{fig:fold10}
\end{figure}

\clearpage
\section{Cross Validation Site Distribution}\label{appendix:cv}

{\small  

\begin{table}[pos= H]
    \centering
    \caption{Number of training and validation sites corresponding to the spatial fold split.}
    \begin{tabular}{lll}
        \toprule
        \textbf{Fold} & \textbf{Train} & \textbf{Validation} \\
        \midrule
        1  & 139 & 13 \\
        2  & 131 & 21 \\
        3  & 137 & 15 \\
        4  & 138 & 14 \\
        5  & 140 & 12 \\
        6  & 139 & 13 \\
        7  & 138 & 14 \\
        8  & 134 & 18 \\
        9  & 135 & 17 \\
        10 & 137 & 15 \\
        \bottomrule
    \end{tabular}
    \label{tab:spatial_folds}
\end{table}

\begin{figure}[pos=h]   
    \centering
    \includegraphics[width=0.5\textwidth]{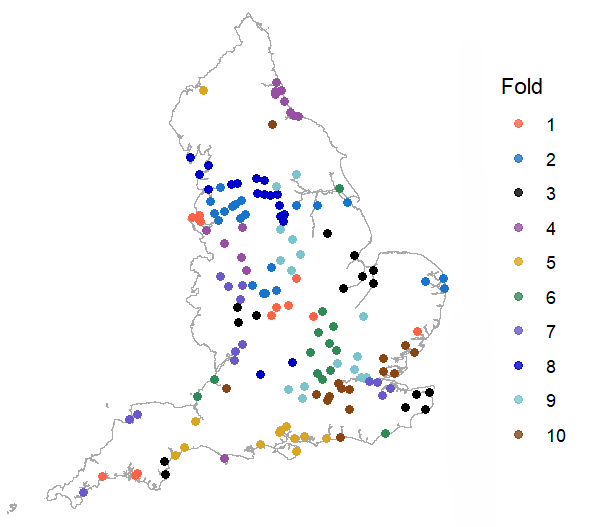}
    \caption{Validation sites by fold across England.}
    \label{fig:validation_site}
\end{figure}

\end{document}